\documentclass[aps,prd,superscriptaddress,nofootinbib, preprintnumbers]{revtex4}
\usepackage{graphicx}
\usepackage{amssymb}
\usepackage{amsmath}
\usepackage{lscape}
\usepackage{textcomp}
\usepackage{epsfig}

\begin{document}

\begin{flushright}
SINP/TNP/2012/08
\end{flushright}
\title{Effect of topological defects and Coulomb charge on the low energy quantum dynamics of gapped graphene}

\author{Baishali Chakraborty}
\email{baishali.chakraborty@saha.ac.in}
\author{Kumar S. Gupta}
\email{kumars.gupta@saha.ac.in}
\affiliation{Theory Division, Saha Institute of Nuclear Physics, 1/AF Bidhannagar, Calcutta 700064, India}
\author{Siddhartha Sen}
\email{siddhartha.sen@tcd.ie}
\affiliation{CRANN, Trinity College Dublin, Dublin 2, Ireland}

\date{\today}

\newcommand{\be}{\begin{equation}} \newcommand{\ee}{\end{equation}}
\newcommand{\bea}{\begin{eqnarray}}\newcommand{\eea}{\end{eqnarray}}

\begin{abstract}
We study the combined effect of a conical topological defect and a Coulomb 
charge impurity on the dynamics of Dirac fermions in gapped graphene. 
Beyond a certain strength of the Coulomb charge, quantum instability sets in, 
which demarcates the boundary between sub and supercritical values of the charge. 
In the subcritical regime, for certain values of the system parameters, the allowed 
boundary conditions in gapped graphene cone can be classified in terms of a single 
real parameter. We show that the observables such as local density of states, 
scattering phase shifts and the bound state spectra are sensitive to the value of this 
real parameter, which is interesting from an empirical point of view. For a supercritical 
Coulomb charge, we analyze the system with a regularized potential as well as with a 
zigzag boundary condition and find the effect of the sample topology on the observable features of the system.
\end{abstract}

\maketitle
\section{Introduction}

The dynamics of Dirac fermions in a 2+1 dimensional conical space-time \cite{Jackiw1}
or in the presence of a cosmic string \cite{gerbert} exhibits a variety of rich nonperturbative 
quantum features. In spite of strong theoretical interest, the quantum properties of
such 2D fermionic systems are difficult to observe in the laboratory.  The experimental 
fabrication of monolayer graphene in 2004 \cite{novo1,novo2,zhang}, whose low energy excitations 
behave like negatively charged fermions satisfying a two dimensional 
Dirac equation \cite{wall,mele,sem,geim,rmp1,rmp2,rmp3}, offers new possibilities to study 
the effect of topological defects in such lower dimensional fermionic 
systems \cite{voz1,voz2,crespi1,crespi2,osi1,sitenko,voz3,voz4,mudry,stone1,furtado,stone2,voz5,Gonzalez,yazyev,fonseca,voz6,voz7,Guinea1,furtado2}. 
The Dirac type excitations in pristine graphene are gapless. However, various impurities, electron-electron interactions,
substrate structures and other short distance effects can violate the sublattice symmetry in graphene, 
leading to a mass gap which has attracted both theoretical \cite{miransky1,khveshchenko1,kane,novikov1,miransky2,khveshchenko2,gusynin,Li,araki1,araki2,zhu}
and experimental \cite{han,zhou1,zhou2,haberer,cooper,andrei2,morpurgo} attention. Thus, the 
gapped graphene system provides a unique template to study the nonperturbative quantum features 
of massive Dirac fermions in the presence of a topological defect.

In graphene the Fermi velocity $v_F \approx10^6 m/s $, which is approximately 300 times 
smaller than the velocity of light. Consequently, a relatively small external Coulomb charge 
impurity $Ze \sim 1$ leads to strong nonperturbative electric field effects 
in graphene \cite{greiner,QED,kats2,novikov1,bridge,castro,levi1,levi2,castro2,kats1,us1,gamayun,milstein,wang,gamayun2,Guinea2}. 
In a gapped graphene system, the external Coulomb charge is said to reach the critical value 
when the system dives into the negative energy continuum \cite{castro2,gamayun,gamayun2} and quantum instability sets in.
Any given external charge in gapped graphene can therefore be classified as either sub or super 
critical. These two different regimes are characterized by markedly different behaviour of the 
observables such as the local density of states (LDOS) \cite{castro2}.

In this paper we shall study the combined effect of a conical topological defect and an external 
Coulomb charge impurity on the low energy quantum dynamics of quasiparticles in gapped graphene. 
When a cone is formed from a graphene sheet, the topological defect introduced in the system gives 
rise to some nontrivial holonomies \cite{crespi1,crespi2,stone1}. The boundary conditions associated 
with the holonomies can be realized  by introducing a suitable flux tube, analogous to a cosmic string, 
passing through the origin \cite{Jackiw1,gerbert,yam,Jackiw2,Jackiw3,Jackiw4,Jackiw5}. In our analysis, 
such a flux tube shall be used to model the conical topological defect on the 2D graphene sheet. 
Let us now consider the effect of an external Coulomb charge impurity in such a system, whose strength could be 
either subcritical or supercritical. For a subcritical Coulomb charge impurity in the presence of the flux tube, we shall show that the 
quantization of the gapped graphene system is not unique and an additional parameter is required to 
fully characterize the boundary conditions at the origin.  In order to understand the physical meaning of such a boundary condition, 
recall that the Dirac description in graphene is valid for low energy or long wavelength excitations. On the other hand, the topological defect as 
well as the Coulomb charge can lead to additional short range interactions, which cannot be incorporated as dynamical terms in the 
Dirac equation. The combined effect of the short range interactions due to the topological defect and the Coulomb charge impurity 
can however be encoded in the boundary conditions\cite{reed,falomir,us2,ksg1}. If we further impose the natural requirement that the graphene system conserves 
probability and the time evolution is unitary, then all the allowed boundary conditions can be labelled by a single real parameter. 
This leads to a one parameter quantization of the gapped graphene system, analogous to what was obtained for Dirac fermions 
in 2+1 dimensional gravity with a topological defect \cite{Jackiw1,gerbert}. For the gapped graphene system, we show that the 
experimental observables such as the LDOS, phase shifts and the bound state energies depend explicitly on the new parameter 
that labels the allowed boundary conditions.

For a supercritical value of the charge impurity in the presence of the topological defect, we study the system 
with a regularized Coulomb potential and also with a zigzag edge boundary condition. The regularization of the 
Coulomb potential takes care of the finite size of the external charge impurity and allows the bound states of 
the system to dive into the negative energy continuum\cite{castro2,gamayun}. The critical charge in gapped graphene cone is renormalized 
to a value higher than that of the gapless case and the value depends on the gap, the cut off parameter, the topology of the system and also 
on the boundary conditions used to obtain the quasibound state spectra in the supercritical region. It will be shown that with the increase 
in gap or cut off parameter the critical charge in presence of zigzag edge boundary condition increases more 
rapidly than in presence of a regularized Coulomb potential.

This paper is organized as follows. In the next Section we set up 
the Dirac equation for gapped graphene cone with a point charge at the 
apex. This is followed by the 
analysis of the spectrum in the subcritical region, 
where we obtain the scattering phase shifts, bound state energies 
and local density of states (LDOS) and show how these physical quantities 
depend explicitly on the sample topology. Then we discuss the effect of 
generalized boundary conditions on the spectrum. In the next section the analysis of the 
corresponding spectrum is done in the supercritical region with a regularized Coulomb 
potential and with zigzag edge boundary condition. 
We end this paper with some discussion and outlook. 
\section{Dirac equation for a gapped graphene cone with a Coulomb charge}

Graphene has a hexagonal honeycomb lattice structure which is formed by 
two inter penetrating triangular sublattices \cite{novo2, zhang, wall, sem, geim}
$A$ and $B$. 
Assuming only nearest neighbour 
hopping in graphene and parameterizing the energy difference between the sublattices 
by $\varepsilon$ we have the Hamiltonian as \cite{sem}
$$H=\beta \sum_{\vec{R_{A}},{i}}[U_{A}^{\dagger}(\vec{R_A})U_{B}(\vec{R_A}+\vec{u_i})+U_{B}^{\dagger}(\vec{R_A}+\vec{u_i})U_{A}(\vec{R_A})]\nonumber$$
\begin{equation}
\label{tba1}
+\varepsilon{\sum}_{\vec{R_A}}[ U_{A}^{\dagger}(\vec{R_A})U_{A}(\vec{R_A})-U_{B}^{\dagger}(\vec{R_A}+\vec{u_1})U_{B}(\vec{R_A}+\vec{u_1})].
\end{equation}
Here $U_{A}^{\dagger}$ and $U_{A}$ ($U_{B}^{\dagger}$ and $U_B$) are the creation and 
destruction operators for electrons localized on sites $A (B)$ respectively. 
The vectors $\vec{u_i}(i=1,2,3)$ connect one $A$ sublattice point to
its three neighbouring $B$ sublattice points.
The hopping parameter $\beta$ is related to the probability amplitude for
electron transfer between neighbouring sites\cite{wall,mele,sem,geim,rmp1,rmp2,rmp3}.
Though for an ideal single layer graphene $\varepsilon=0$, 
by breaking the sublattice symmetry a gap can be introduced in graphene\cite{miransky1,khveshchenko1,kane,novikov1,miransky2,
khveshchenko2,gusynin,Li,araki1,araki2,zhu,han,zhou1,zhou2,haberer,cooper,andrei2,morpurgo} 
and in our following work we shall consider the massive Dirac excitations of a gapped graphene cone. 

\begin{figure}
\centering
\begin{tabular}{cc}
\epsfig{file=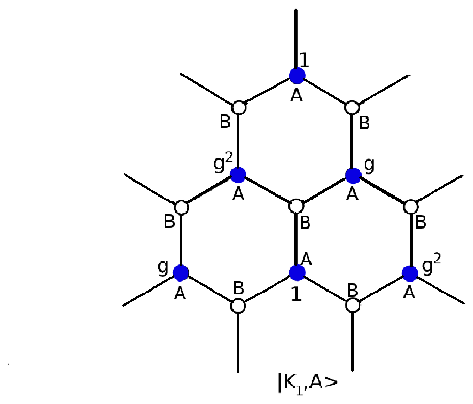,width=0.5\linewidth,clip=} &
\epsfig{file=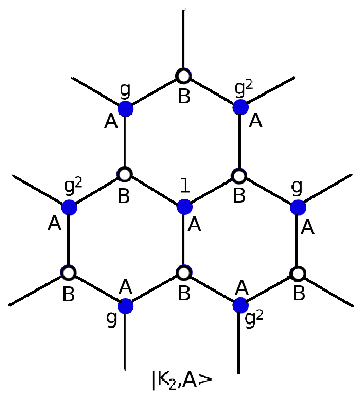,width=0.5\linewidth,clip=}\\
\epsfig{file=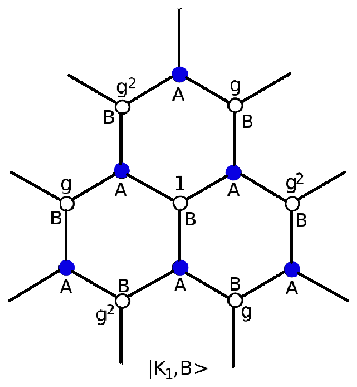,width=0.5\linewidth,clip=} &
\epsfig{file=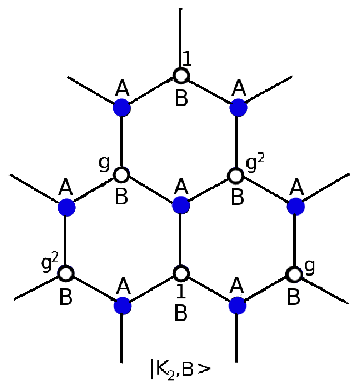,width=0.5\linewidth,clip=}\\
\end{tabular}
\caption{ The linearly independent energy eigenstates of graphene are shown.
Here the solid and empty circles belong to sublattice A and B respectively. $1$, $g$ and $g^2$ 
represent the nonzero amplitudes of the wavefunction at the lattice sites where they are assigned
and the wavefunction has zero amplitude at all the other remaining lattice sites.
Here $g= \mbox{exp}(i 2\pi/3)$ and $g^2= \mbox{exp}(-i 2\pi/3)$. }
\label{fig:eigenstates}
\end{figure}

From the Hamiltonian $H$ we obtain that in gapped graphene the energy eigenvalues are minimum at the six 
vertices of the first Brillouin zone of graphene and they are known as the Dirac points. Among these points, 
two are inequivalent \cite{novo2, zhang, wall, sem, geim}. We consider them to be
situated at the opposite corners of the Brillouin zone and
we denote their wave vectors by $\mathbf{K}_1$ and $\mathbf{K}_2$. 
Thus we can construct the four linearly independent energy eigenstates \cite{wall, geim, rmp1}
of the hopping Hamiltonian denoted by $|K_1,A\rangle,|K_1,B\rangle,|K_2,A\rangle$ 
and $|K_2,B\rangle$.
The pseudospin indices $A$ and $B$ in the eigenstates correspond to that sublattice on which the 
wavefunction has nonzero amplitude and the valley indices $K_1$ and $K_2$ in 
the eigenstates are distinguished by the manner in which 
the phase of the wavefunction evolves around a lattice site having zero amplitude wavefunction \cite{stone2} (see Fig.\ref{fig:eigenstates}).
It can also be seen from Fig.\ref{fig:eigenstates} that the states with valley index $K_2$ can be produced by rotating the corresponding
states with valley index $K_1$ by $180^{\circ}$\cite{crespi2}.
The basis is chosen in such a manner that $\mathbf{K}_2=-\mathbf{K}_1$.
The low energy eigenstates in graphene can be expressed as a linear combination 
of these energy eigenstates multiplied by envelope functions varying slowly 
on the lattice parameter scale.

The low-energy properties of the quasiparticle states in graphene near the Dirac point 
having valley index $K_1$, can be described by the Dirac equation 
\begin{equation}
\label{h1.1}
 H \Psi = \left [ -i(\sigma_1 \partial_x + \sigma_2 \partial_y) + m\sigma_3 \right ] \Psi = E \Psi, 
\end{equation} 
where $m$ denotes the Dirac mass generated due to sublattice symmetry breaking, $E$ is the energy eigenvalue and we have set $\hbar = v_F = 1$. The Hamiltonian acts on the array of the slowly varying envelope functions
$\Psi= \left( 
\begin{array}{c}
 \Psi_{K_1,A} \\
 \Psi_{K_1,B} 
\end{array}
\right)$
 The Pauli matrices $\sigma_{1,2,3}$ act on the pseudospin indices $A,B$. 

\begin{figure}
\centering
\begin{tabular}{cc}
\epsfig{file=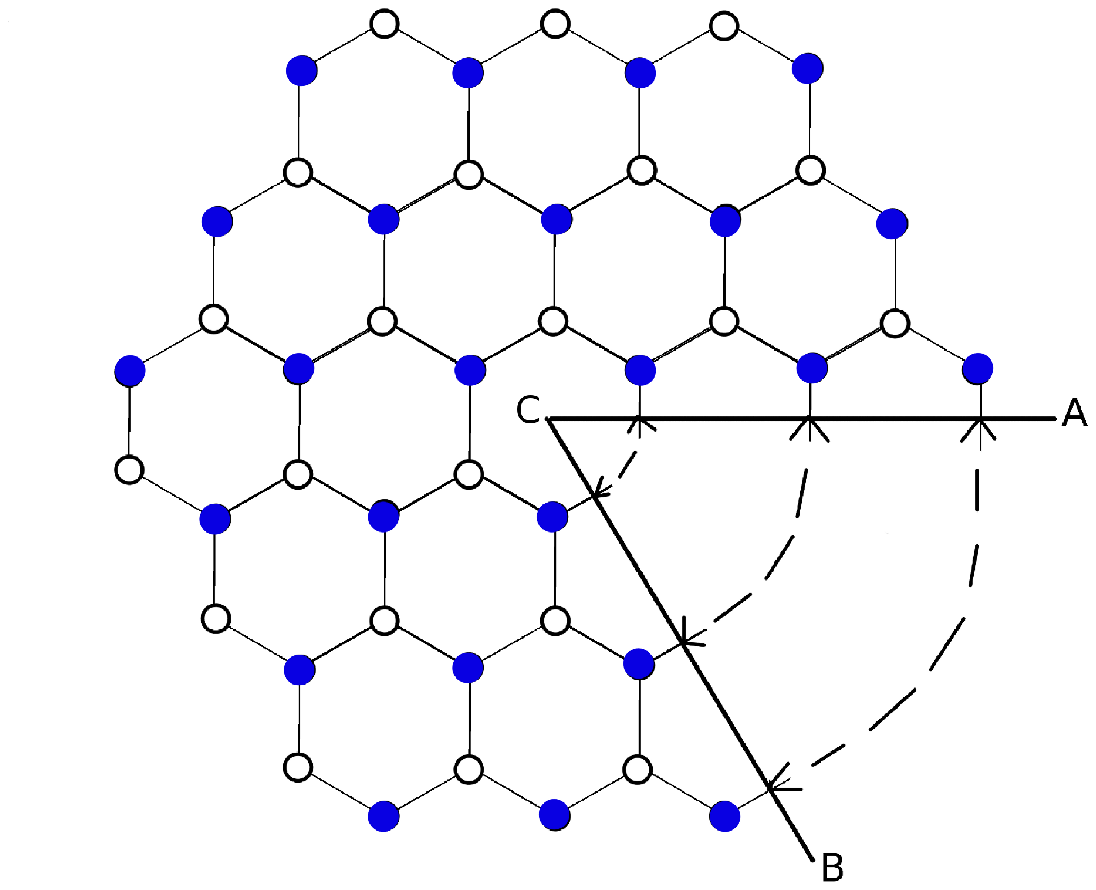,width=0.5\linewidth,clip=} &
\epsfig{file=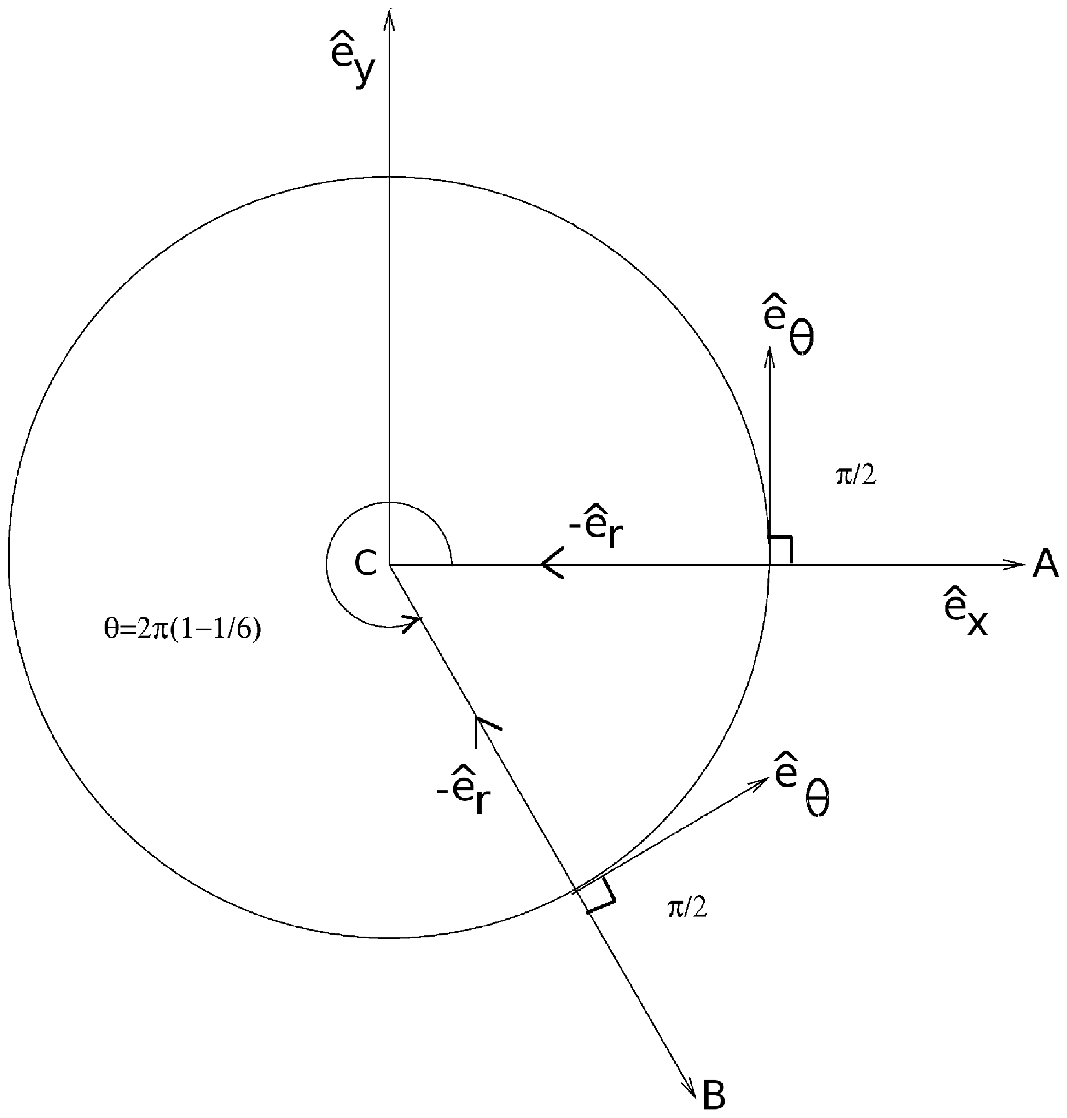,width=0.5\linewidth,clip=}\\
\end{tabular}
\caption{ Formation of cone from plane graphene sheet and rotation of the coordinate system in order to make it continuous.}
\label{fig:1}
\end{figure}


To study the effect of topology on this system, the formation of a graphene cone is considered 
by introducing local defects in the hexagonal lattice structure 
of graphene\cite{crespi1,crespi2}. When a sector is removed from the 
plane sheet of graphene and the two edges of the sector are identified,
the frame $\{\hat{e}_x,\hat{e}_y\}$ becomes discontinuous across the joining line.
Therefore we choose a new set of frames $\{\hat{e}_{x^\prime},\hat{e}_{y^\prime}\}$
which is rotated with respect to the old frame by an angle 
$\varphi =\theta + \frac{\pi}{2}$ in the counter clockwise direction (see Fig.\ref{fig:1}).
The $x^\prime$ and $y^\prime$ axes are chosen along the $\hat{e}_\theta$ direction
and the $-\hat{e}_r$ direction respectively \cite{crespi1,crespi2}.


For this change of reference frame the wave function has to be transformed by $\mbox{exp}(i\varphi\sigma_3/2)$
to keep the form of the Hamiltonian the same\cite{crespi1,crespi2}.
Thus the conical topology gives rise to nontrivial 
holonomies for the pseudoparticle wavefunctions. When a cone with angle of deficit 
$\frac{2n\pi}{6}$ is formed, where $n$ can take only discrete values $1,2,3,4,5$, 
the angular boundary condition obeyed by the Dirac spinor as it goes around a closed path is given by 
\begin{eqnarray}
\label{h1.6}
\Psi(r,\theta=2\pi)=e^{i2\pi(1-\frac{n}{6}) \frac{\sigma_3}{2}}\Psi(r,\theta=0).
\end{eqnarray} 
Here $(r,\theta)$ denotes the polar coordinate of the lattice points.

When the cone is formed by removing odd number of wedges of angle $\frac{2\pi}{6}$
from the plane graphene sheet and the two edges of the removed portion are identified, 
the adjacent sites on two sides of the identification line belong to the 
same sublattice (see Fig.\ref{fig:1}). Thus the bipartite nature of the hexagonal lattice
is broken. Also from Fig.\ref{fig:eigenstates} we can see that rotation of a state with valley index $K_1$
by an odd multiple of  angle $\frac{2\pi}{6}$ gives the corresponding state with valley index $K_2$
with the same sublattice label.
Therefore the removal of odd number of wedges of angle $\frac{2\pi}{6}$ gives rise to an additional 
phase shift affecting the valley indices of the wave function in the boundary condition\cite{crespi1,crespi2,osi1,stone2}.
The states with valley index $K_{2}$ will be affected in the same manner as the states with valley index $K_{1}$
but there will be a relative phase difference of 180$^\circ$ between
them. Therefore this boundary condition can be described by involving  
a $\tau_{2}$ matrix in it where the matrix $\tau_2$ operate on the valley indices\cite{crespi1,crespi2,critical}.
When $n$ is even, this off diagonal matrix does not play 
any role and the exponential factor appearing in the boundary condition
just gives $\pm1$ depending on the value of $n$.
We diagonalize the matrix $\tau_{2}$ for all allowed odd values of $n$. 
As a result the valley indices of the electronic states become mixtures of $K_{1}$ and $K_{2}$.
Then the angular boundary condition satisfied for all values of $n$, by a branch of electronic states having
a fixed Fermi index, is given by \cite{crespi1,crespi2}
\begin{eqnarray}
\label{h1.9}
\Psi(r,\theta=2\pi)=e^{i2\pi[\pm \frac{n}{4} \sigma_0 +(1-\frac{n}{6}) \frac{\sigma_3}{2}]}\Psi(r,\theta=0).
\end{eqnarray}
 Here $\sigma_0$ is an identity matrix which acts on the pseudospin indices $A,B$.
$\Psi= \left( 
\begin{array}{c}
 \Psi_{A,K^{\prime}} \\
 \Psi_{B,K^{\prime}} 
\end{array}
\right)$ where $K^{\prime}$ is a mixture of $K_{1}$ and $K_{2}$.

The effect of these  holonomies can be modelled by introducing
a fictitious magnetic flux tube \cite{voz3} passing through the apex of the cone.
The magnetic vector potential modifies the boundary condition on a Dirac spinor as 
\begin{eqnarray}
\label{h5}
\Psi(r,\theta=2\pi)=e^{ie\oint\vec{A}\cdot\vec{dl} }\Psi(r,\theta=0).
\end{eqnarray}
Here $\vec{dl}$ is a line element on the circumference of the cone at a distance $r$ from the apex, i.e.
\begin{eqnarray}
\label{h6}
\vec{dl}= \hat{e}_{\theta}~r(1-\frac{n}{6})d\theta. 
\end{eqnarray}
Substituting (\ref{h6}) in (\ref{h5}) and 
assuming that the component $A_\theta$ of the magnetic vector potential
is independent of the angle $\theta$, we have from Eq.(\ref{h1.9})
\begin{eqnarray}
\label{h8}
A_\theta = \frac{1}{er}[\pm\frac{\frac{n}{4}\sigma_0}{(1-\frac{n}{6})}+\frac{\sigma_3}{2}].
\end{eqnarray}
Then an external Coulomb charge localized at the apex of the gapped graphene cone can be 
equivalently described by a suitable combination of electric charge and magnetic 
flux tube \cite{critical}.
Let us assume that the Coulomb interaction strength is
$\alpha = \frac{Ze^2}{\kappa}$, where $Z$ is the atomic number 
of the impurity, $e$ is the electronic charge and $\kappa$ is the dielectric constant.
Replacing the ordinary derivatives in the Hamiltonian by the corresponding 
covariant derivatives, the Dirac equation for the low energy excitations of 
gapped graphene cone in presence of a Coulomb charge at its apex is given by
\begin{eqnarray} 
\label{h11}
H \Psi(r,\theta) = \left( 
\begin{array}{cc}
m-\frac{\alpha}{r} &  \partial_r - \frac{i}{r(1-\frac{n}{6})}\partial_\theta \pm \frac{\frac{n}{4}}{r(1-\frac{n}{6})} + \frac{1}{2r}  \\
-\partial_r - \frac{i}{r(1-\frac{n}{6})}\partial_\theta  \pm \frac{\frac{n}{4}}{r(1-\frac{n}{6})}-\frac{1}{2r} & 
-m-\frac{\alpha}{r}
\end{array}
\right)
\left( 
\begin{array}{c}
 \Psi_{A}(r,\theta)\\
 \Psi_{B}(r,\theta)
\end{array}
\right) = E \left( 
\begin{array}{c}
 \Psi_{A}(r,\theta) \\
 \Psi_{B}(r,\theta) 
\end{array}
\right).
\end{eqnarray}
Let
\begin{eqnarray}
\label{h12}
\Psi(r,\theta)=\sum_j \left( 
\begin{array}{c}
\Psi_A^{(j)}(r)\\
\Psi_B^{(j)}(r)
\end{array}
\right)e^{ij\theta},
\end{eqnarray}
where $j$ is half-integer.
Substituting (\ref{h12}) in (\ref{h11}), we obtain that the leading short distance 
behavior of the wavefunction is given by 
\begin{equation}
\label{h16}
\Psi_{A,B}^{(j)}(r)\sim r^{\gamma-\frac{1}{2}}~~~~\mbox{where}~~~~
\gamma = \sqrt{\nu^2 - \alpha^2}~~~~~\mbox{and}~~~~~\nu=\frac{(j\pm\frac{n}{4})}{(1-\frac{n}{6})}.
\end{equation}
We should note that the angular part of the wave function is different from that of the
planer case, due to the choice of the reference frame\cite{crespi2}.
From (\ref{h16}) we can see that when $|\alpha|$ exceeds $|\nu|$, $\gamma$ 
becomes imaginary. Therefore, the eigenstates $\Psi_A^{(j)}(r)$ 
and $\Psi_B^{(j)}(r)$ becomes wildly oscillatory and have no well 
defined limit as $r\rightarrow0$. For massive excitations the critical coupling $\alpha_c$ corresponds
to that value of $\alpha$ for which $E=-m$. When $m=0$, the value of $\alpha_c$ is equal to the minimum 
allowed value of $\nu$ and depending on the magnitude of Dirac mass and boundary conditions $\alpha_c$
increases gradually from $\nu$. It will be shown 
that the critical coupling for the gapped graphene cone explicitly depends on the 
angle of the cone and also on the product of gap and cutoff parameter. 
From the expression of $\nu$ one can see that if we consider the expression
$\nu=\frac{(j-\frac{n}{4})}{(1-\frac{n}{6})}$ and  
$j=\frac{3}{2}$, then $\nu = \frac{3}{2}$ for all values of $n$. Therefore
for analyzing the effect of topology the angular 
momentum channel $j=\frac{3}{2}$ has not been considered.

Depending on the strength of the external Coulomb charge compared to that 
of the critical charge of a gapped graphene cone with a particular 
opening angle, the effect of the charge impurity on the cone can be analyzed
in two separate regions: subcritical and supercritical.

\section{Dirac equation for a gapped graphene cone with a subcritical Coulomb charge}

In this Section we discuss the bound and scattering state solutions of the Dirac fermions 
in a gapped graphene cone in the presence of an external Coulomb charge impurity. Following \cite{novikov1}, consider the ansatz
\begin{equation}
\label{sub1.1}
 \Psi_A^j(\rho)=\sqrt{m+E}e^{-\frac{\rho}{2}} {\rho}^{\gamma - \frac{1}{2}}[F(\rho) + G(\rho) ]
\end{equation}
and
\begin{equation}
\label{sub1.2}
 \Psi_B^j(\rho)=\sqrt{m-E}e^{-\frac{\rho}{2}} {\rho}^{\gamma - \frac{1}{2}}[F(\rho) - G(\rho) ],
\end{equation}
where $\rho=2 \eta r ,~~\eta=\sqrt{m^2-E^2},~~\gamma=\sqrt{\nu^2 - \alpha^2},~~\nu=\frac{(j\pm\frac{n}{4})}{(1-\frac{n}{6})}$ and
total angular momentum $j$ takes all half integer values. Using Eqs. (\ref{h11}), (\ref{h12}), (\ref{sub1.1}) and (\ref{sub1.2}) we get
\begin{equation} 
\label{sub1.3}
H_\rho \left( \begin{array}{c}
 {F(\rho)} \\
 {G(\rho)} \\ 
\end{array} \right)=
\left( 
\begin{array}{cc}
\rho \frac{d}{d \rho} + \left(\gamma -\frac {\alpha E}{\eta}\right) & - \left(\nu +\frac{m\alpha}{\eta}\right)  \\
\left(-\nu + \frac{m\alpha}{\eta}\right) & \rho \frac{d}{d \rho} + \left(\gamma -\rho +\frac {\alpha E}{\eta}\right)
\end{array}
\right)\left( \begin{array}{c}
 {F(\rho)} \\
 {G(\rho)} \\
\end{array} \right)=0,
\end{equation}
where $H_\rho$ denotes the radial Dirac operator.
From Eq.(\ref{sub1.3}) we have
\begin{equation} \label{sub2}
 \rho \frac{d F}{d \rho} + \left(\gamma -\frac {\alpha E}{\eta}\right) F - \left(\nu + \frac{m\alpha}{\eta}\right) G = 0.~~~~~~~~~
\end{equation}
and
\begin{equation} \label{sub3}
 \rho \frac{d G}{d \rho} + \left(\gamma -\rho +\frac {\alpha E}{\eta}\right) G + \left(-\nu + \frac{m\alpha}{\eta}\right) F = 0.~~~~~~~~~
\end{equation}
Substituting the expression of $G$ from Eq.(\ref{sub2}) in Eq.(\ref{sub3}) we have
\begin{equation}\label{sub4}
 \rho F^{\prime\prime} + (1+2\gamma-\rho)F^{\prime} - \left(\gamma-\frac{\alpha E}{\eta}\right)F=0.
\end{equation}

In order to proceed, for the moment we assume that the wavefunction vanishes at the charge impurity. 
Solutions of Eq. (\ref{sub4}) which obey that boundary condition are given by \cite{stegun}
\begin{equation}\label{sub8}
 F(\rho)= A_1 M\left(\gamma -\frac{E\alpha}{\eta}, 1+2\gamma, \rho\right),
\end{equation}
where $A_1$ is a constant. 
From Eq.(\ref{sub2}) we have
\begin{equation}\label{sub9}
 G(\rho)= \frac{\left(\gamma-\frac{\alpha E}{\eta}\right)}{\left(\nu +\frac{m\alpha}{\eta}\right)}A_1 M\left(1+\gamma -\frac{E\alpha}{\eta}, 1+2\gamma, \rho\right).
\end{equation}
The upper and lower components of the wavefunctions are
\begin{equation}\label{sub10}
 \Psi_A^j(\rho)=A_1 \sqrt{m+E} e^{-\frac{\rho}{2}}\rho^{\gamma-\frac{1}{2}}\left[M\left(\gamma -\frac{E\alpha}{\eta}, 1+2\gamma, \rho\right)+\frac{\left(\gamma-\frac{\alpha E}{\eta}\right)}{\left(\nu +\frac{m\alpha}{\eta}\right)}M\left(1+\gamma -\frac{E\alpha}{\eta}, 1+2\gamma, \rho\right)\right]
\end{equation}
and
\begin{equation}\label{sub11}
\Psi_B^j(\rho)=A_1 \sqrt{m-E} e^{-\frac{\rho}{2}}\rho^{\gamma-\frac{1}{2}}\left[M\left(\gamma -\frac{E\alpha}{\eta}, 1+2\gamma, \rho\right)-\frac{\left(\gamma-\frac{\alpha E}{\eta}\right)}{\left(\nu +\frac{m\alpha}{\eta}\right)}M\left(1+\gamma -\frac{E\alpha}{\eta}, 1+2\gamma, \rho\right)\right]. 
\end{equation}
Bound states occur when the wavefunctions reduce to polynomials i.e. when
\begin{equation}\label{sub12}
 \gamma-\frac{\alpha E}{\eta}=-p,
\end{equation}
where
\begin{eqnarray}\label{sub13}
p=\left\{\begin{array}{l}
0,1,2,..., \ \mbox{when $\nu>0$}, \\
1,2,3...., \ \mbox{when $\nu<0$}.
\end{array}\right.
\end{eqnarray}
The corresponding bound state spectra is obtained as
\begin{eqnarray}\label{sub14}
E_p=\frac{m \ \mbox{sgn}(\alpha)}{\sqrt{1+\frac{{\alpha}^2}{(p+\gamma)^2}}}.
\end{eqnarray}
Here the energy should be of the same sign (positive or negative) as $\alpha$ 
because otherwise the value of $p$ will become negative and 
in our range of interest, it is not allowed.
 
The solution of Eq.(\ref{sub1.3}) which leads to physical scattering 
states when $ |E| > |m|$ is \cite{stegun}
\begin{equation}\label{sub5}
 F(\rho)= A_1 M\left(\gamma -\frac{E\alpha}{\eta}, 1+2\gamma, \rho\right) + A_2 \rho^{-2\gamma}M\left(-\gamma-\frac{E\alpha}{\eta}, 1-2\gamma, \rho\right).
\end{equation}
From Eq.(\ref{sub2}) we have
\begin{equation}\label{sub6}
 G(\rho)= \frac{\left(\gamma-\frac{\alpha E}{\eta}\right)}{\left(\nu +\frac{m\alpha}{\eta}\right)}A_1 M\left(1+\gamma -\frac{E\alpha}{\eta}, 1+2\gamma, \rho\right)-\frac{\left(\gamma+\frac{\alpha E}{\eta}\right)}{\left(\nu +\frac{m\alpha}{\eta}\right)}A_2 \rho^{-2\gamma}M\left(1-\gamma-\frac{E\alpha}{\eta}, 1-2\gamma, \rho\right).
\end{equation}
Here the parameter $ \; \eta = \sqrt{m^{2} - E^{2} } \; $ is 
purely imaginary, i.e.  $ \; \eta = -ik, \;$\cite{novikov1} where $k$ is defined as $k= \sqrt{E^{2} - m^{2}} $. Consequently, 
the variable $ \; \rho \;$ also becomes purely imaginary, 
$ \; \rho = -2 ik r.$ Using the $r \rightarrow\infty$ limit of the 
scattering states the scattering matrix is obtained as
\begin{equation}\label{sub15}
S(k)=(2ik)^{\frac{2i\alpha E}{k}}\frac{\left(\nu+i\frac{m \alpha}{k}\right)}{\left(\gamma-i\frac{E\alpha}{k}\right)}\frac{\Gamma\left(1+\gamma-i\frac{\alpha E}{k}\right)}{\Gamma\left(1+\gamma+i\frac{E\alpha}{k}\right)}e^{i\pi\left(\gamma +i\frac{\alpha E}{k}\right)}. 
\end{equation}
From Eq.(\ref{sub15}) it can be seen that the poles of the $S$ matrix determined by  
$\left(1+\gamma-i\frac{\alpha E}{k}\right)=1-p$, where $p$ is a nonzero positive integer and 
$\left(\gamma-i\frac{E\alpha}{k}\right)=0$ when $\nu>0$, gives back the 
corresponding bound states as expected.

\subsection{Generalized boundary conditions}

The Dirac equation discussed in the previous section is valid
for low energy or long wavelength excitations. The conical defect as well as the Coulomb charge impurity might give rise to 
short range interactions in the system, which cannot be incorporated as dynamical terms in the Dirac equation.
However, the combined effect of those short range interactions can be taken into account through the choice of suitable 
boundary conditions. In systems with unitary time evolution, there is a well defined prescription due to von Neumann to 
determine the allowed boundary conditions, which is what we shall follow \cite{reed,falomir,us2,ksg1}.  

From Eq.(\ref{h11}) it can be seen that the Dirac operator $H$ has an 
angular part and a radial part. The angular part operates on a domain $Y(\theta)$
which is spanned by the antiperiodic functions $e^{ij\theta}$ where $j$ is 
a half integer and the corresponding boundary condition is kept unchanged.
The radial Dirac operator $H_{\rho}$, given by Eq.(\ref{sub1.3}), is symmetric 
in the domain $\mathcal{D}_{0} = C_{0}^{\infty}(R^{+})$ which consists of infinitely differentiable 
functions of compact support in the real half line $R^{+}$ and its adjoint 
operator $H_{\rho}^{\dagger}$ has the same expression as $H_{\rho}$ but its domain can be different.
Now to determine the domain of self-adjointness of the Dirac operator $H$, consider the equations 
\begin{equation}\label{sub16}
 H^{\dagger}\Psi_{\pm}=\pm\frac{i}{d}\Psi_{\pm},
\end{equation}
where $d$ has the dimension of length. 
The total number of square integrable, linearly independent solutions of Eq.(\ref{sub16})
gives the deficiency indices for $H$ and they are denoted by $n_{\pm}$.
For obtaining $n_{\pm}$, Eq.(\ref{sub1.3}) is considered with $E$ replaced by $\pm\frac{i}{d}$.
To understand the significance of these indices we should note that if an operator is 
self-adjoint, then it is expected to have only real eigenvalues. Thus the existence of 
imaginary eigenvalues $\pm\frac{i}{d}$ in the spectrum is a measure of the deviation of 
an operator from self-adjointness. The non zero deficiency indices serve as the measurement of this deviation.
Depending on the deficiency indices $H_\rho$ can be classified in three different ways \cite{reed} :
$(1)$ When $n_+ = n_- =0$, $H_{\rho}$ is essentially self-adjoint in $\mathcal{D}_{0}(H_{\rho})$. 
$(2)$ When $n_+ = n_-\neq0$, $H_{\rho}$ is not self-adjoint in $\mathcal{D}_{0}(H_{\rho})$ but it 
can admit self-adjoint extensions. 
$(3)$ When $n_+\neq n_-$, $H_{\rho}$ cannot have self-adjoint extensions. 

To find the deficiency indices $n_\pm$ let us first consider the following.
\begin{equation}
\label{sub16.1}
 \Psi_{A\pm}^j(\rho)=\sqrt{m\pm\frac{i}{d}}e^{-\frac{\rho}{2}} {\rho}^{\gamma - \frac{1}{2}}[F_{\pm}(\rho) + G_{\pm}(\rho) ]
\end{equation}
and
\begin{equation}
\label{sub16.2}
 \Psi_{B\pm}^j(\rho)=\sqrt{m\mp \frac{i}{d}}e^{-\frac{\rho}{2}} {\rho}^{\gamma - \frac{1}{2}}[F_{\pm}(\rho) - G_{\pm}(\rho) ],
\end{equation}
where $\rho=2 \eta_1 r ,~~\eta_1=\sqrt{m^2 + \frac{1}{d^2}},~~\gamma=\sqrt{\nu^2 - \alpha^2}$ and $\nu=\frac{(j\pm\frac{n}{4})}{(1-\frac{n}{6})}$.
Then we can write
\begin{equation} 
\label{sub16.3}
H_\rho \left( \begin{array}{c}
 {F_{\pm}(\rho)} \\
 {G_{\pm}(\rho)} \\ 
\end{array} \right)=
\left( 
\begin{array}{cc}
\rho \frac{d}{d \rho} + \left(\gamma \mp\frac {i\alpha}{\eta_1 d}\right) & - \left(\nu +\frac{m\alpha}{\eta_1}\right)  \\
\left(-\nu + \frac{m\alpha}{\eta_1}\right) & \rho \frac{d}{d \rho} + \left(\gamma -\rho \pm\frac {i\alpha}{\eta_1 d}\right)
\end{array}
\right)\left( \begin{array}{c}
 {F_{\pm}(\rho)} \\
 {G_{\pm}(\rho)} \\
\end{array} \right)=0.
\end{equation}
From Eq.(\ref{sub16.3}) we have
\begin{equation} 
\label{sub17}
 \rho \frac{d F_{\pm}(\rho)}{d \rho} + \left(\gamma \mp \frac {i\alpha }{\eta_1 d}\right) F_{\pm}(\rho) - \left(\nu +\frac{m\alpha}{\eta_1}\right) G_{\pm}(\rho) = 0,
\end{equation}
and
\begin{equation} \label{sub18}
 \rho \frac{d G_{\pm}(\rho)}{d \rho} + \left(\gamma -\rho \pm \frac {i\alpha}{\eta_1 d}\right) G_{\pm}(\rho) + \left(-\nu + \frac{m\alpha}{\eta_1}\right) F_{\pm}(\rho) = 0.~~~~~~~~~
\end{equation}
Substituting the expression of $G_{\pm}(\rho)$ from Eq.(\ref{sub17}) in Eq.(\ref{sub18}) we have
\begin{equation}\label{sub19}
 \rho F_{\pm}^{\prime\prime}(\rho) + (1+2\gamma-\rho)F_{\pm}^{\prime}(\rho) - \left(\gamma\mp\frac{i\alpha}{\eta_1 d}\right)F_{\pm}(\rho)=0.
\end{equation}
We first determine the deficiency subspace characterized by $F_{+}(\rho)$ and $G_{+}(\rho)$
given in Eq.(\ref{sub16.3}). The required solution of Eq.(\ref{sub19}) is
\begin{equation}\label{sub20}
 F_+ (\rho)= U\left(\gamma-\frac{i\alpha}{\eta_1 d},1+2\gamma,\rho\right).
\end{equation}
Using the differential recursive relation
   $\; z U'(a,b,z) + a U(a,b,z) = a(1+a-b) U(a + 1,b,z),  \; $ from Eq.(\ref{sub17}) we have 
\begin{equation}\label{sub21}
 G_+ (\rho)= \frac{\left(\gamma-\frac{i\alpha}{\eta_1 d}\right)\left(-\gamma-\frac{i\alpha}{\eta_1 d}\right)}{\left(\nu+\frac{m\alpha}{\eta_1}\right)}U\left(1+\gamma-\frac{i\alpha}{\eta_1 d},1+2\gamma,\rho\right).
\end{equation}
In the limit $\rho \longrightarrow 0$ the functions behave as
\begin{eqnarray} \label{a6}
F_+ &\longrightarrow& \frac{\pi}{\sin \pi (1+ 2 \gamma)}(P_+ - Q_+ \rho^{-2 \gamma}), \\
G_+ &\longrightarrow& \frac{\pi}{\sin \pi (1+ 2 \gamma)} (R_+ - S_+ \rho^{-2 \gamma}),
\end{eqnarray}
where 
\begin{eqnarray} \label{a7}
P_+ &=& \frac{1}{\Gamma(-\gamma - \frac{i \alpha}{\eta_1 d}) \Gamma(1 + 2 \gamma)}\\ 
Q_+ &=& \frac{1}{\Gamma(\gamma - \frac{i \alpha}{\eta_1 d}) \Gamma(1 - 2 \gamma)} \\
R_+ &=& \frac{( \gamma - \frac{i \alpha}{\eta_1 d} )
 (- \gamma - \frac{i \alpha}{\eta_1 d})}{( \nu + \frac{m \alpha}{\eta_1})}
   \frac{1}{\Gamma(1 -\gamma - \frac{i \alpha}{\eta_1 d}) \Gamma(1 + 2 \gamma)} \\
S_+ &=& \frac{( \gamma - \frac{i \alpha}{\eta_1 d} )
 (- \gamma - \frac{i \alpha}{\eta_1 d})}{( \nu + \frac{m \alpha}{\eta_1})}
 \frac{1}{\Gamma(1 + \gamma - \frac{i \alpha}{\eta_1 d}) \Gamma(1 - 2 \gamma)} 
\end{eqnarray}
are constants depending on the system parameters. 
From the above relations we find that as $\rho \longrightarrow 0$,
\begin{equation} \label{gb1}
\int {|\psi_{A+}|}^2 \rho d\rho \longrightarrow \int (a_1 \rho^{2 \gamma} \ +a_2 \ + \ a_3 \rho^{-2 \gamma}) d\rho,
\end{equation} 
\begin{equation} \label{gb2}
\int {|\psi_{B+}|}^2 \rho d\rho \longrightarrow \int (b_1 \rho^{2 \gamma} \ +b_2 \ + \ b_3 \rho^{-2 \gamma}) d\rho,
\end{equation} 
where $a_i$,$b_i$ $(i=1,2,3)$ are constants, whose explicit forms 
are not relevant. As $\gamma$ is a real positive quantity 
in the subcritical region, from Eq.(\ref{gb1}) and (\ref{gb2}) it can be 
shown that $\psi_{A+}$ and $\psi_{B+}$ are square 
integrable everywhere provided $0 < \gamma < \frac{1}{2}$. 
Thus $n_+=1$ for the parameter range $0 < \gamma < \frac{1}{2}$.

In a similar way, by analyzing the deficiency subspace 
characterized by the negative sign, we obtain
\begin{equation} \label{a7.1}
 F_- = U \left ( \gamma + \frac{i \alpha}{\eta_1 d}, 1 + 2 \gamma, \rho \right ), 
\end{equation}
\begin{equation} \label{a7.2}
 G_- = \frac{\left ( \gamma + \frac{i \alpha}{\eta_1 d} \right ) 
    \left (- \gamma + \frac{i \alpha}{\eta_1 d} \right )}{\left ( \nu + \frac{m \alpha}{\eta_1} \right )} 
    U \left (1+ \gamma + \frac{i \alpha}{\eta_1 d}, 1 + 2 \gamma, \rho \right ). 
\end{equation}
In addition, in the limit $\rho \longrightarrow 0$ the functions $F_-$ and $G_-$ behave as
\begin{eqnarray} \label{a7.3}
F_- &\longrightarrow& \frac{\pi}{\sin \pi (1+ 2 \gamma)} (P_- - Q_- \rho^{-2 \gamma}), \\
G_- &\longrightarrow& \frac{\pi}{\sin \pi (1+ 2 \gamma)} (R_- - S_- \rho^{-2 \gamma}),
\end{eqnarray}
where 
\begin{equation} \label{a7.5}
P_- = {\bar{P}}_+, \;\;\;\; Q_- = {\bar{Q}}_+, \;\;\;\;
R_- = {\bar{R}}_+, \;\;\;\; S_- = {\bar{S}}_+ .\;\;\;\; 
\end{equation}
Similar analysis as before shows that $n_-=1$ for the parameter range 
$0 < \gamma < \frac{1}{2}$ as well. Thus for the gapped
graphene cone with a charge impurity, $n_+ = n_- =1$ when $0 < \gamma < \frac{1}{2}$. 
Therefore, this system admits a one parameter
family of self-adjoint extensions for $0 < \gamma < \frac{1}{2}$.
We would now like to find out the spectrum of the system in a range 
of $\nu$ and the effective subcritical Coulomb potential strength 
$\alpha$ such that $0<\gamma<\frac{1}{2}$. The deficiency subspaces 
for the radial Dirac operator $H_{\rho}$ are spanned by 
the elements,
\begin{eqnarray}\label{gbb1}
\Psi_{\pm} \ = \left( \begin{array}{c}
 {{\Psi}_{A \pm} } \\
 {{\Psi}_{B \pm} }\\
\end{array} \right) \ = \left( \begin{array}{c}
 {\sqrt{m \pm \frac{i}{d}} e^{-\frac{\rho}{2}} \rho^{\gamma -\frac{1}{2}} (F_\pm + G_\pm )} \\
 {\sqrt{m \mp \frac{i}{d}} e^{-\frac{\rho}{2}} \rho^{\gamma -\frac{1}{2}} (F_\pm - G_\pm )}\\
\end{array} \right).
\end{eqnarray}
The domain in which the Dirac operator is self-adjoint is given by 
${\mathcal{D}}_{z}(H_\rho) = {\mathcal{D}}_{0}(H_{\rho})\oplus  \{C(e^{i \frac{z}{2}} {\Psi}_{+} + e^{-i\frac{z}{2}}{\Psi}_{-})
\},$ where $C$ is an arbitrary complex number 
and $z\in R ~\mbox{mod} ~2\pi$. Thus we have a one parameter family of 
self-adjoint extensions, labeled by a real parameter $z$. 
For each choice of the parameter $z$, we have a domain of 
self-adjointness of the radial Dirac operator defined by 
${\mathcal{D}}_{z}(H_\rho)$. When $\rho \longrightarrow 0$ an 
arbitrary element $\Psi_z \in {\mathcal{D}}_{z}(H_\rho)$ can be written as
\begin{eqnarray}\label{gbb2}
\Psi_{z} \ = \left( \begin{array}{c}
 {{\Psi}_{Az} } \\
 {{\Psi}_{Bz} }\\
\end{array} \right) \ \longrightarrow \ C\left( \begin{array}{c}
{\sqrt{m + \frac{i}{d}} e^{\frac{iz}{2}} \rho^{\gamma -\frac{1}{2}} (F_{+} + G_{+}) \ + \ \sqrt{m-\frac{i}{d}} e^{-\frac{iz}{2}} \rho^{\gamma -\frac{1}{2}} (F_{-} + G_{-}}) \\
{\sqrt{m-\frac{i}{d}} e^{\frac{iz}{2}} \rho^{\gamma -\frac{1}{2}} (F_{+} - G_{+}) \ + \ \sqrt{m + \frac{i}{d}} e^{-\frac{iz}{2}} \rho^{\gamma -\frac{1}{2}} (F_{-} - G_{-}})\\
\end{array} \right),~~~~~
\end{eqnarray}
 where $F_{-} ~\mbox{and} ~G_{-}$ denote the complex conjugates of $F_{+} ~\mbox{and} ~G_{+}$ respectively.

Now the spectrum of the system is found out when the boundary 
conditions are governed by the domain $\mathcal D_z(H_\rho).$ 
A solution of the physical eigenvalue problem is written as
\begin{eqnarray}\label{gbb3}
\Psi= N\left( \begin{array}{c}
 {\sqrt{m + E} e^{-\frac{\rho}{2}} \rho^{\gamma -\frac{1}{2}} (F(\rho) + G(\rho) )} \\
 {\sqrt{m - E} e^{-\frac{\rho}{2}} \rho^{\gamma -\frac{1}{2}} (F(\rho) - G(\rho) )}\\
\end{array} \right)
\end{eqnarray}
Here $F$ and $G$ satisfy Eqs. (\ref{sub2}) and (\ref{sub3}) respectively, 
and $N$ denotes the normalization. Solutions of 
Eqs.(\ref{sub2}) and (\ref{sub3}) that are square integrable at infinity
are given by
\begin{equation} \label{gbb4}
F = U \left ( \gamma - \frac{\alpha E}{\eta}, 1 + 2 \gamma, \rho \right ), 
\end{equation}
\begin{equation} \label{gbb5}
G = U \left (1+ \gamma - \frac{\alpha E}{\eta}, 1 + 2 \gamma, \rho \right ).
\end{equation}
Now using Eqs.(\ref{gbb2}) and (\ref{gbb3}) we have in the limit $\rho \longrightarrow 0$,
\begin{eqnarray} \label{gbb6}
F \longrightarrow \frac{\pi}{\sin \pi (1+ 2 \gamma)} (P - Q \rho^{-2 \gamma}), \\
G \longrightarrow \frac{\pi}{\sin \pi (1+ 2 \gamma)} (R - S \rho^{-2 \gamma}),
\end{eqnarray}
 where
\begin{eqnarray} \label{gbb7}
P = \frac{1}{\Gamma\left(-\gamma - \frac{\alpha E}{\eta}\right) \Gamma\left(1 + 2 \gamma\right)} ~~~~~~~
Q = \frac{1}{\Gamma\left(\gamma - \frac{\alpha E}{\eta}\right) \Gamma\left(1 - 2 \gamma\right)} ~~~~~~\\
R = \frac{1}{\Gamma\left(1 -\gamma - \frac{\alpha E}{\eta}\right) \Gamma\left(1 + 2 \gamma\right)} ~~~~
S = \frac{1}{\Gamma\left(1 + \gamma - \frac{\alpha E}{\eta}\right) \Gamma\left(1 - 2 \gamma\right)}. 
\end{eqnarray}
Hence, as $\rho \longrightarrow 0$,
\begin{eqnarray}\label{gbb8}
 \Psi \longrightarrow \ \frac{\pi}{\sin \pi (1+ 2 \gamma)}N\left( \begin{array}{c}
{\sqrt{m + E}[(P+R) \rho^{\gamma -(\frac{1}{2})} - (Q+S)\rho^{-\gamma -(\frac{1}{2})}]} \\
{\sqrt{m - E}[(P-R) \rho^{\gamma -(\frac{1}{2})} - (Q-S)\rho^{-\gamma -(\frac{1}{2})}]}\\
\end{array} \right)
\end{eqnarray}
The physical solution $\Psi$ in Eq.(\ref{gbb8}) must belong to the domain of self-adjointness given by 
$\mathcal D_z(H_\rho)$. In fact behavior of the elements of the domain $\mathcal D_z(H_\rho)$ determines the boundary 
conditions for the system. If $\Psi_z \in \mathcal D_z(H_\rho)$, then as $\rho \longrightarrow 0$ the coefficients of 
$r^{\gamma-(1/2)}$ and $r^{-\gamma-(1/2)}$ in Eqs. (\ref{gbb2}) and (\ref{gbb8}) must match.
Let us define 
$$\sqrt{m + \frac{i}{d}}(P_{+}+R_{+})=\chi_1 e^{i\phi_1}$$
$$\mbox{and} \quad \sqrt{m + \frac{i}{d}}(Q_{+}+S_{+})=\chi_2 e^{i\phi_2} .~~~~~~~~~$$
\begin{figure}
\centering
\begin{tabular}{cc}
\epsfig{file=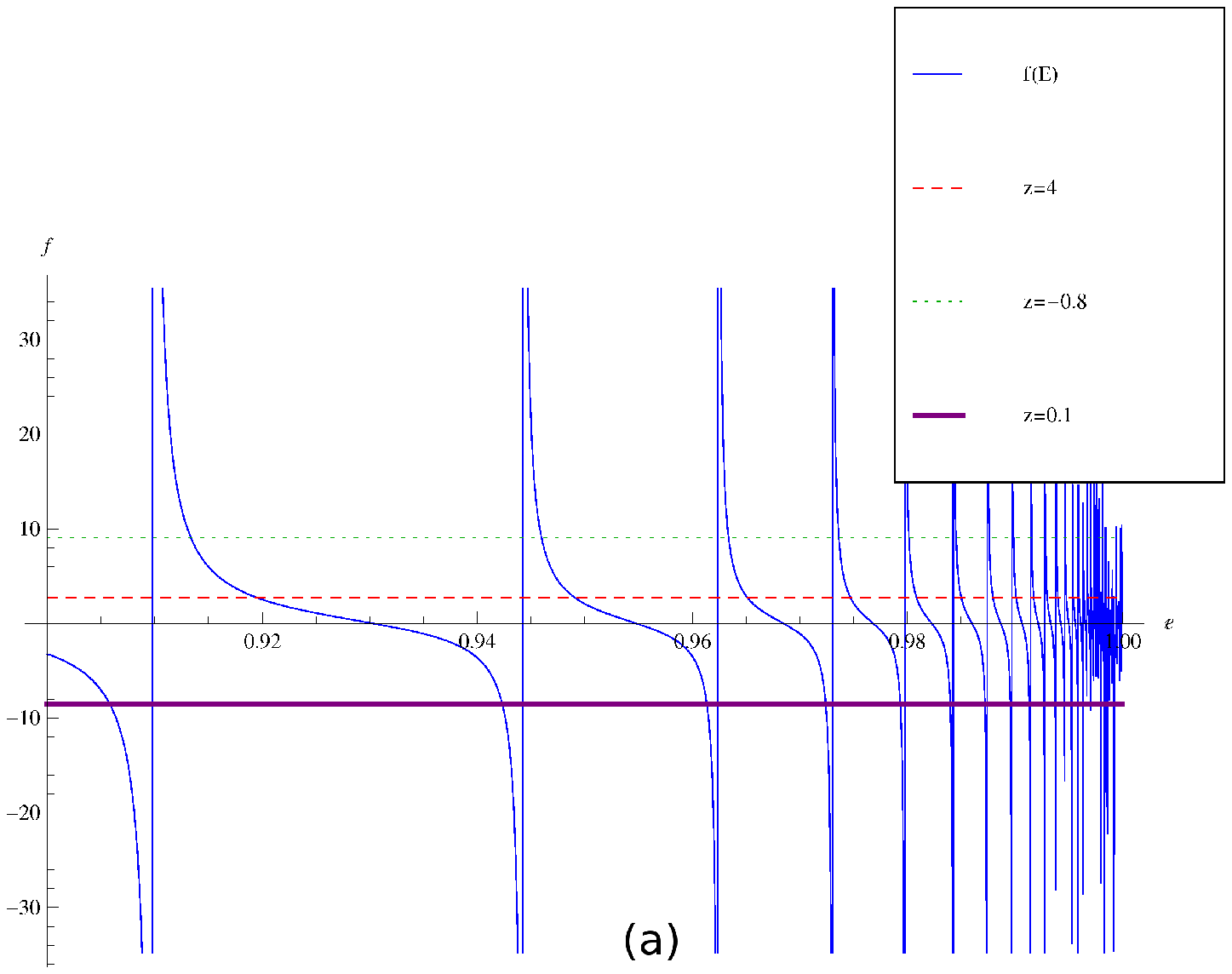,width=0.5\linewidth,clip=} &
\epsfig{file=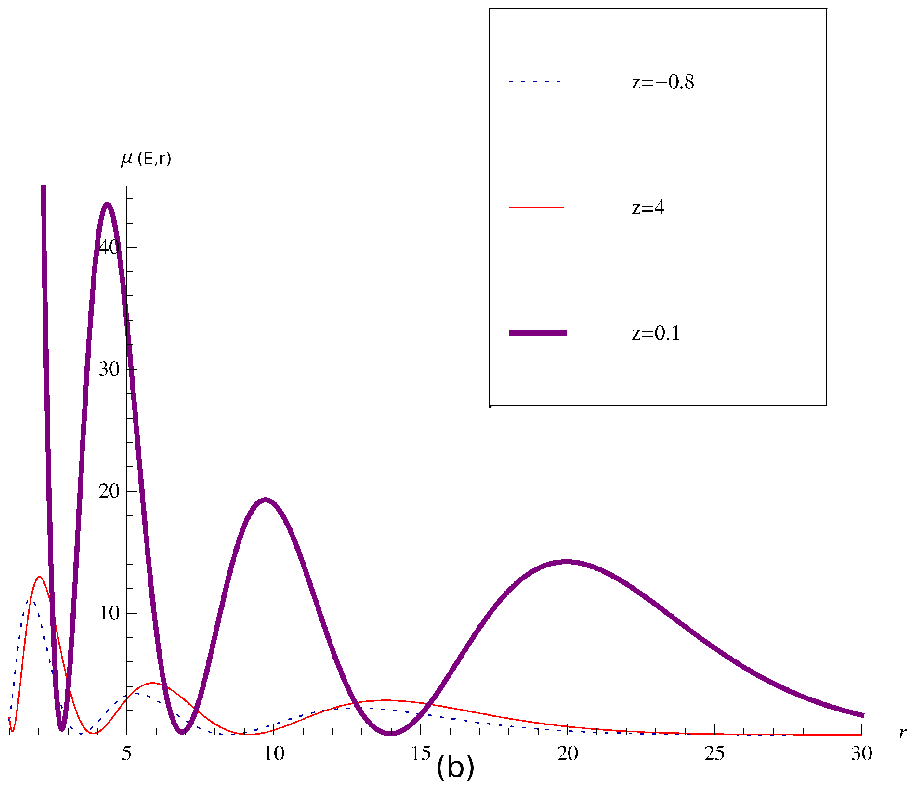,width=0.5\linewidth,clip=}\\
\epsfig{file=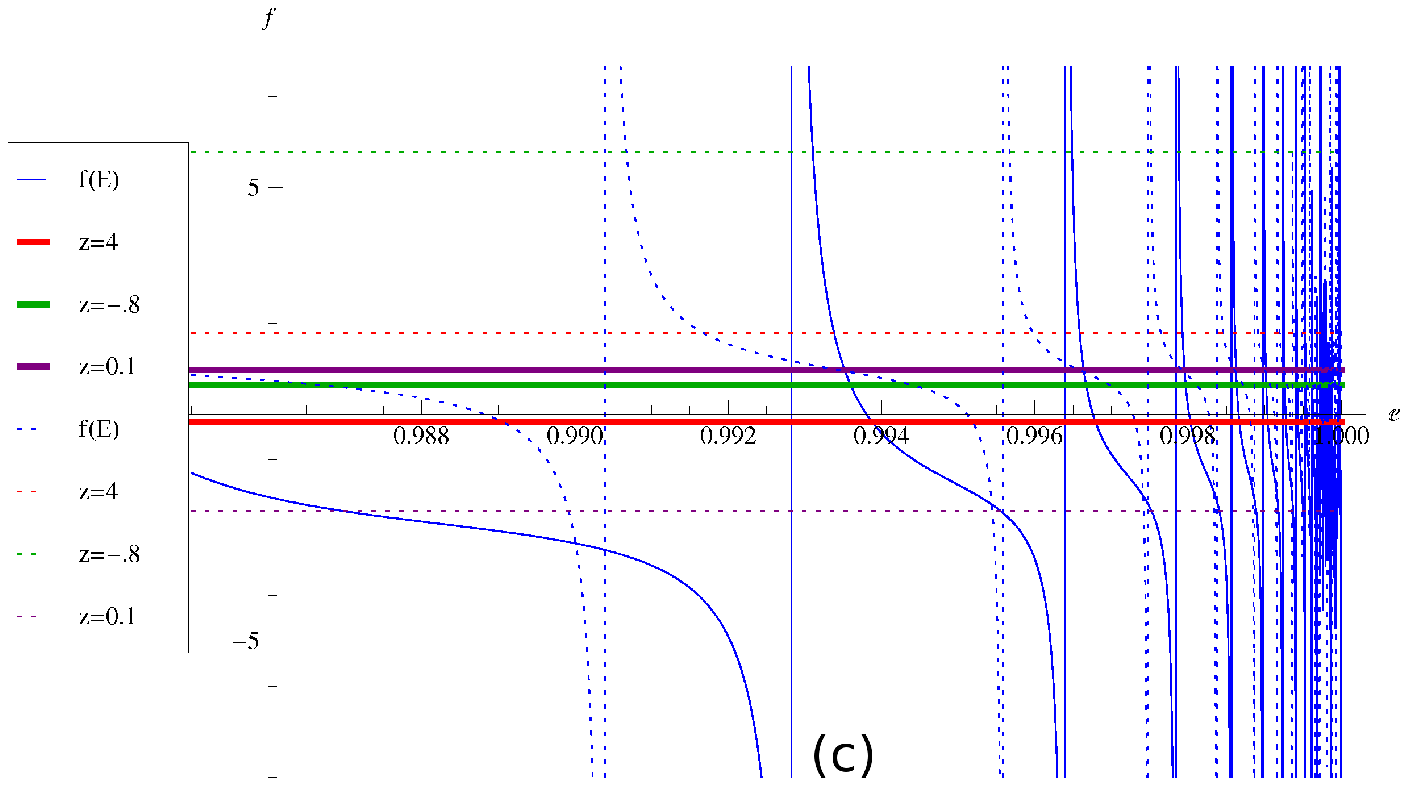,width=0.5\linewidth,clip=} &
\epsfig{file=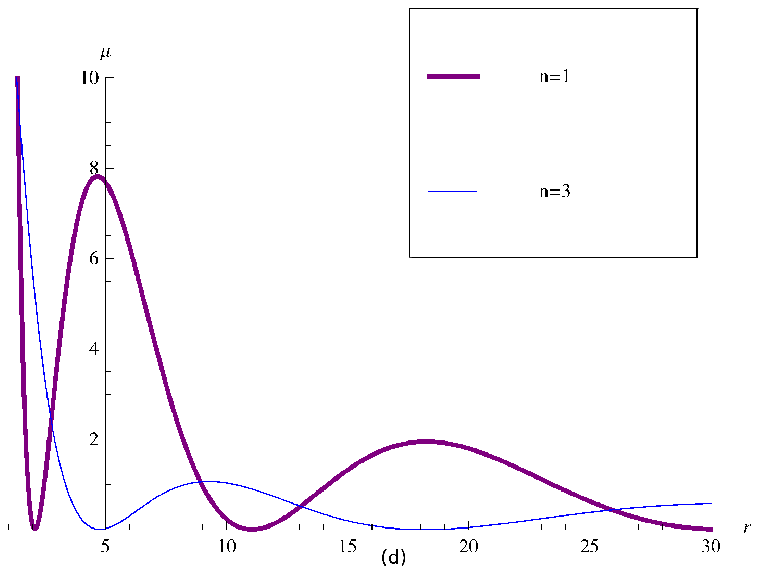,width=0.5\linewidth,clip=}\\
\end{tabular}
\caption{(a)Plot of $f(E)$ is shown for system parameters
$j=\frac{3}{2},~n=1,~\alpha=1.48$ and $m=1$. The three horizontal
line corresponds to the three different values of the self adjoint 
extension parameter $z=4,0.1,-0.8$.
(b) Dependence of LDOS in the bound state sector of the gapped graphene
cone on the distance $r$ from the external charge impurity is shown for three
different values of bound state energy corresponding to three different values of
self adjoint extension parameter. Here the contribution from angular momentum 
channel $j=\frac{3}{2}$ is shown and the system parameters are $n=1,~\alpha=1.48$ and $m=1$.
We have assumed d=1.
(c) Plot of $f(E)$ is shown for two different values of $n$ ($1$ and $3$) with system parameters
$j=\frac{1}{2}$, $\alpha=0.29$ and $m=1$. The three horizontal
line corresponds to the three different values of the self adjoint 
extension parameter $z=4,0.1,-0.8$. The solid lines correspond to $n=3$ and the dotted lines 
correspond to $n= 1$.
(d) Dependence of LDOS on the distance $r$ from the external charge impurity is shown for
two different values of bound state energy corresponding to two different values of $n$ 
with the self-adjoint extension parameter $z=0.1$. Here the contribution from angular momentum 
channel $j=\frac{1}{2}$ is shown and the system parameters are $\alpha=0.29$ and $m=1$. }
\label{fig:6}
\end{figure}
Now comparing the terms of Eqs. (\ref{gbb2}) and (\ref{gbb8}) we obtain
\begin{eqnarray}\label{gbb9}
\bigg(\frac{\eta^2}{\frac{1}{d^2}+m^2}\bigg)^{\gamma} \frac{P+R}{Q+S}=\frac{\chi_{1} \mbox{cos}\left(\phi_{1} + \frac{z}{2} \right)}{\chi_{2} \mbox{cos}\left(\phi_{2} + \frac{z}{2} \right)}
\end{eqnarray}
Using the expressions of the constants $P,Q,R ~\mbox{and}~ S $ and the above equation we finally get
$$f(E) \equiv \bigg(\frac{\eta^2}{\frac{1}{d^2}+m^2}\bigg)^{\gamma} \frac{\left(1 -\gamma - \frac{\alpha E}{\eta}\right)
\Gamma\left(1 + \gamma - \frac{\alpha E}{\eta}\right)  \Gamma\left(1 - 2 \gamma\right)}{\left(1 +\gamma -\frac{\alpha E}{\eta}\right)
\Gamma\left(1 -\gamma - \frac{\alpha E}{\eta}\right)\Gamma\left(1 + 2 \gamma\right)}$$ 
\begin{eqnarray}\label{gbb10}
=\frac{\chi_{1} \mbox{cos}\left(\phi_{1} + \frac{z}{2} \right)}{\chi_{2} \mbox{cos}\left(\phi_{2} + \frac{z}{2} \right)}.~~~~~~~~~~~~~~~~~~~~~~~~~~~~~~~~~
\end{eqnarray}
Eq.(\ref{gbb10}) determines the spectrum in terms of the system 
parameters and the self-adjoint extension parameter $z$. Each 
choice of $z$ corresponds to a different boundary condition 
described by the domain $\mathcal D_z(H_\rho)$ and leads to an 
inequivalent quantum theory. However the choice of $z$ for a 
particular system is determined empirically as the theory cannot 
predict its value. For a special choice of $z=z_1$ such that 
$\phi_2 + \frac{z_1}{2}=\frac{\pi}{2},$ we have
\begin{eqnarray}
\gamma-\frac{E\alpha}{\eta}=-p, \quad p=1,2,3,.....  .
\end{eqnarray}
This leads to the spectrum obtained in Eq.(\ref{sub12}) for $0<\gamma<\frac{1}{2}$. 
For another special choice of $z=z_2$ such 
that $\phi_1 +\frac{z_2}{2}=\frac{\pi}{2}$, we get
\begin{eqnarray}
-\gamma-\frac{E\alpha}{\eta}=-p, \quad p=1,2,3,... . 
\end{eqnarray}
Though Eq.(\ref{gbb10}) cannot be solved analytically, 
from a typical plot of $f(E)$ it can be obtained numerically. From 
Figure $(3a)$ we can see when $z$ changes from $-0.8$
to $4$, the bound state energy changes from $0.915$ to $0.92$.
Again when $z$ changes from $4$ to $0.1$, the bound state energy 
changes from $0.92$ to $0.942$. Now calculating the contribution
to LDOS coming from a single angular momentum channel $j=\frac{3}{2}$
for these three different values of bound state energy, we observe 
in diagram $3(b)$ how the $r$ dependence of LDOS varies with different 
values of $z$. In Figure $(3c)$ we have shown how the bound
state energy depends on the topology of the system for a particular 
angular momentum channel $j=\frac{1}{2}$ and  three self-adjoint 
extension parameters $z= -0.8,4$ and $0.1$. From the Figure $(3c)$ we can 
see when $n$ changes from $3$ to $1$, for the self-adjoint extension parameter
$z=0.1$ the bound state energy changes from $0.9935$ to $0.9955$. Calculating the 
contribution to LDOS coming from a single angular momentum channel $j=\frac{1}{2}$
for these two different values of bound state energy, we observe 
in diagram $3(d)$ how the topology of a system affects the $r$ dependence of LDOS .

In the scattering state sector where $|E|>|m|$ and $\eta= -ik$ the general solution of 
Eqs.(\ref{sub2}) and (\ref{sub3}) that lead to physical scattering 
states are given by
\begin{equation}\label{scat1}
 F(\rho)= P_1(k) M\left(\gamma -\frac{E\alpha}{\eta}, 1+2\gamma, \rho\right) + Q_1(k) \rho^{-2\gamma}M\left(-\gamma-\frac{E\alpha}{\eta}, 1-2\gamma, \rho\right).
\end{equation}
From Eq.(\ref{sub2}) we have
\begin{equation}\label{scat2}
 G(\rho)= \frac{\left(\gamma-\frac{\alpha E}{\eta}\right)}{\left(\nu +\frac{m\alpha}{\eta}\right)}P_1(k) M\left(1+\gamma -\frac{E\alpha}{\eta}, 1+2\gamma, \rho\right)-\frac{\left(\gamma+\frac{\alpha E}{\eta}\right)}{\left(\nu +\frac{m\alpha}{\eta}\right)}Q_1(k) \rho^{-2\gamma}M\left(1-\gamma-\frac{E\alpha}{\eta}, 1-2\gamma, \rho\right).
\end{equation}
Substituting these expressions of $F$ and $G$ in Eq (\ref{sub1.1}) and 
Eq (\ref{sub1.2}) we get the upper and lower components of the wave function.
Then from the asymptotic form of the wavefunction, identifying the incoming 
and outgoing waves, the scattering matrix and the phase shifts are obtained.
Now, to find a relation between the constants $P_1(k)$ and $Q_1(k)$, we consider
the short distance limit of the wave function.   
\begin{figure}
\centering
\begin{tabular}{cc}
\epsfig{file=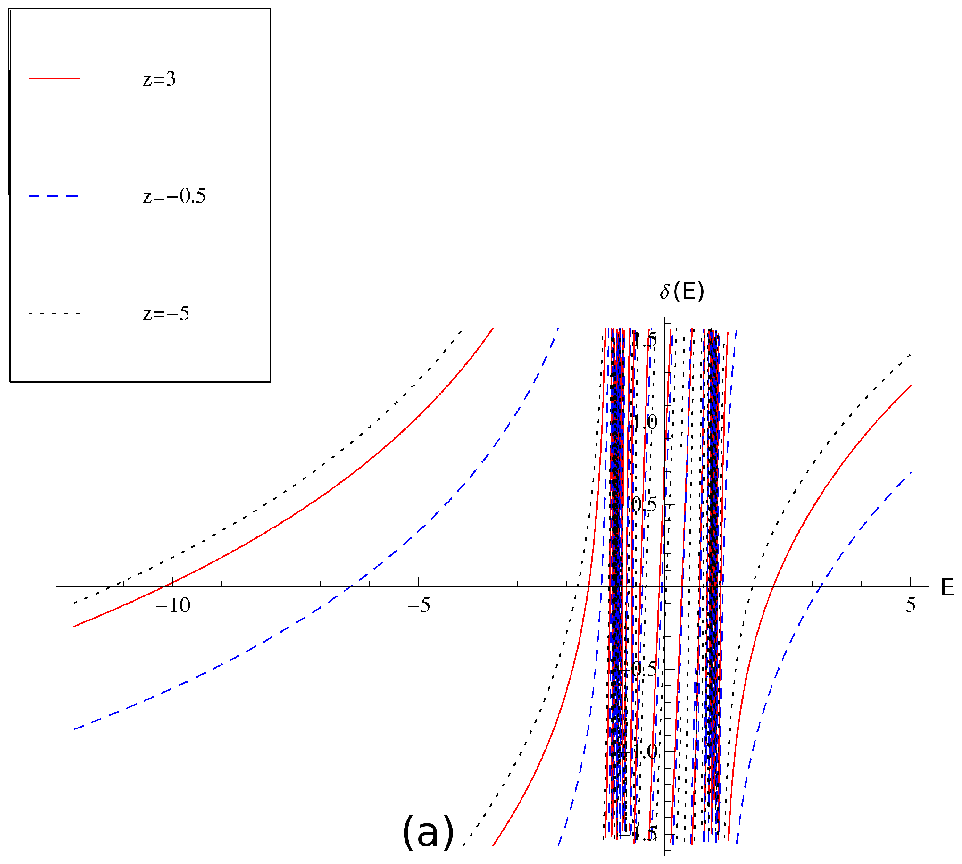,width=0.5\linewidth,clip=} &
\epsfig{file=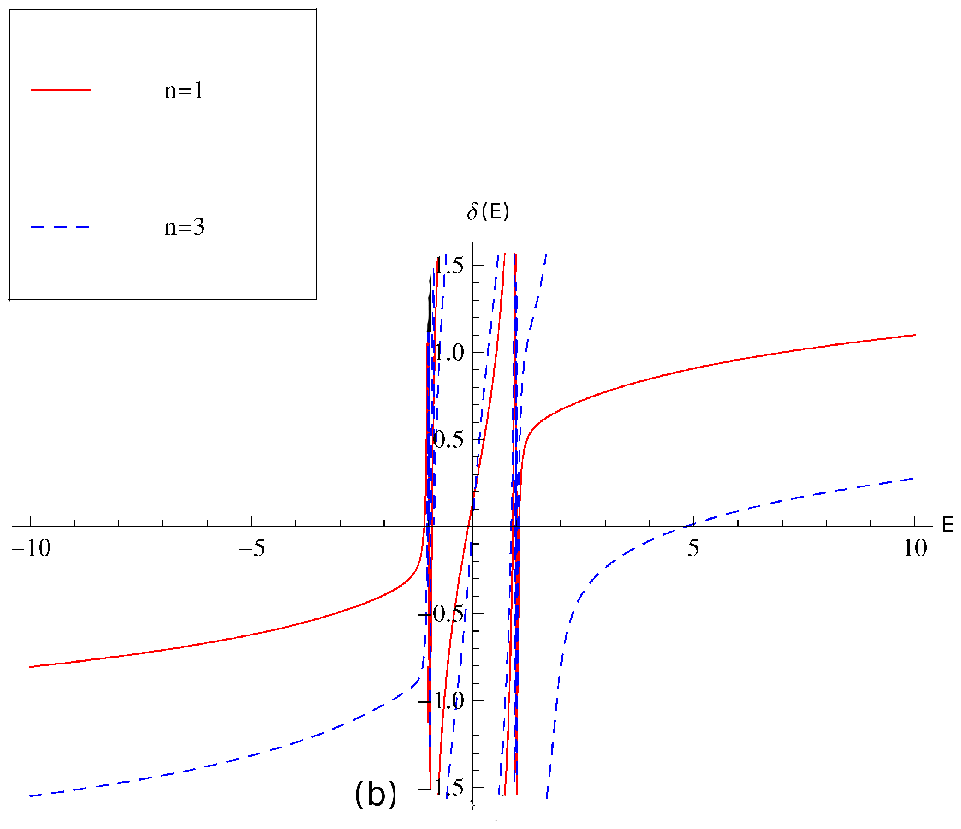,width=0.5\linewidth,clip=}\\
\end{tabular}
\caption{(a)Phase shifts in the gapped graphene cone is shown 
for three different values of the self adjoint extension 
parameter $z=3,-0.5,-5$ where the system parameters are 
$n=1,j=\frac{3}{2},\alpha=1.48,$ and $m=1$.
(b) Scattering phase shifts are shown for different angles of 
the gapped graphene cone with the sae parameter $z=-0.5$ and system 
parameters $j=\frac{1}{2}, \alpha=0.29$ and $m=1$.}
\label{fig:7}
\end{figure}
The domain of self-adjointness of the Hamiltonian $H_{\rho}$ is given by 
 ${\mathcal{D}}_{z}(H_{\rho}) = {\mathcal{D}}(H_{\rho})\oplus  \{e^{i \frac{z}{2}} {\Psi}_{+} + e^{-i\frac{z}{2}}{\Psi}_{-}
\}.$ In the limit $\; r \rightarrow 0 ,\; $  an element of the domain ${\mathcal{D}}_{z}(H_{\rho})$ can be given by
\begin{equation} \label{gbs1}
   \Psi = C \left
    ( e^{i \frac{z}{2}} {\Psi}_{+} + e^{-i \frac{z}{2}} {\Psi}_{-}
    \right ),
\end{equation}
where $C$ is an arbitrary constant.

 \begin{figure}
\centering
\begin{tabular}{cc}
\epsfig{file=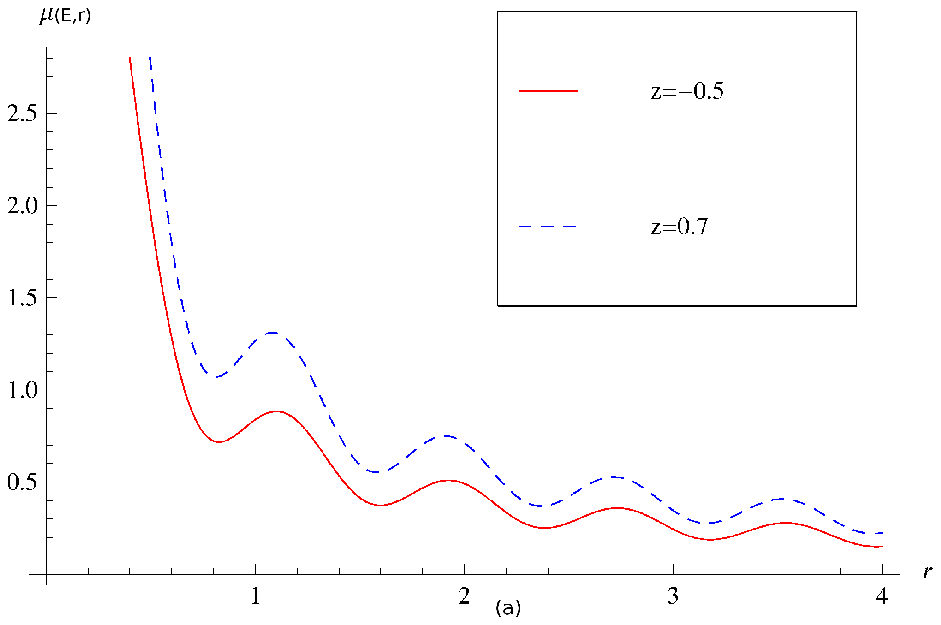,width=0.5\linewidth,clip=} &
\epsfig{file=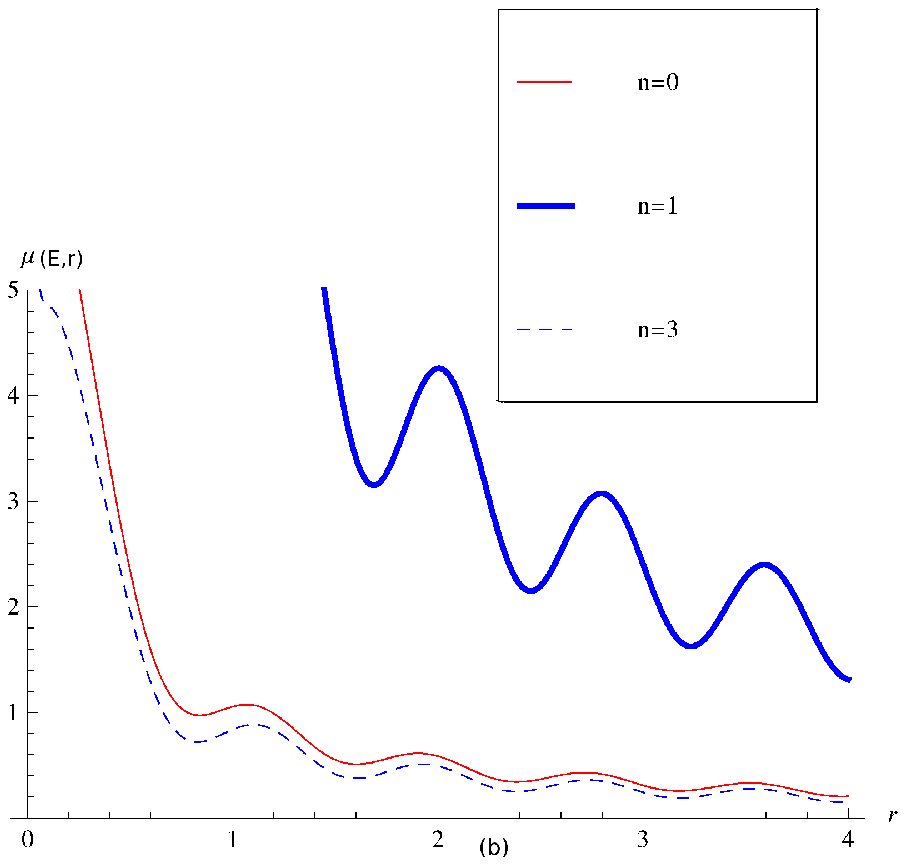,width=0.5\linewidth,clip=}\\
\epsfig{file=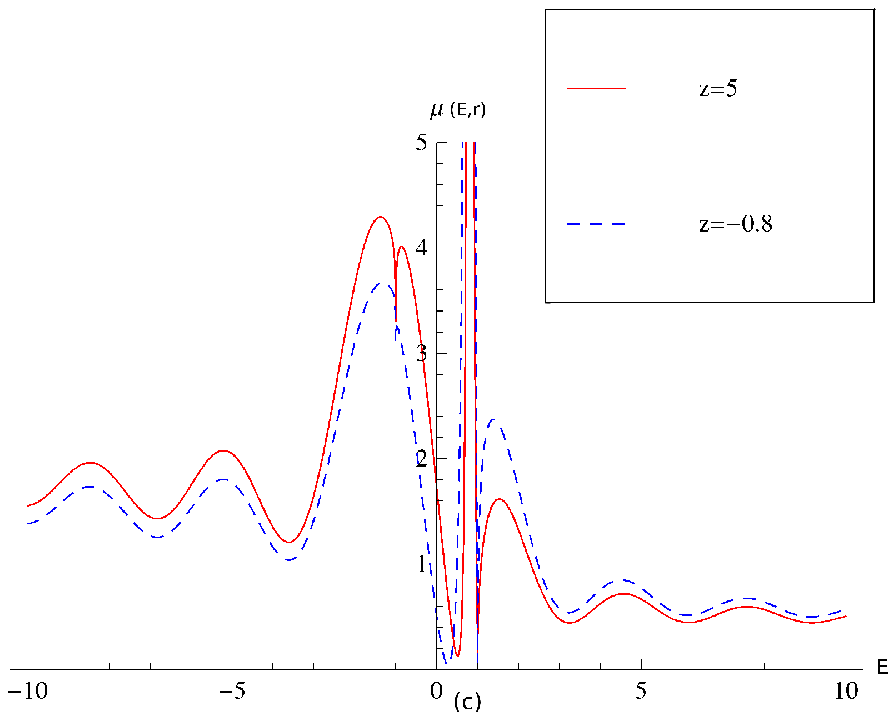,width=0.5\linewidth,clip=} &
\epsfig{file=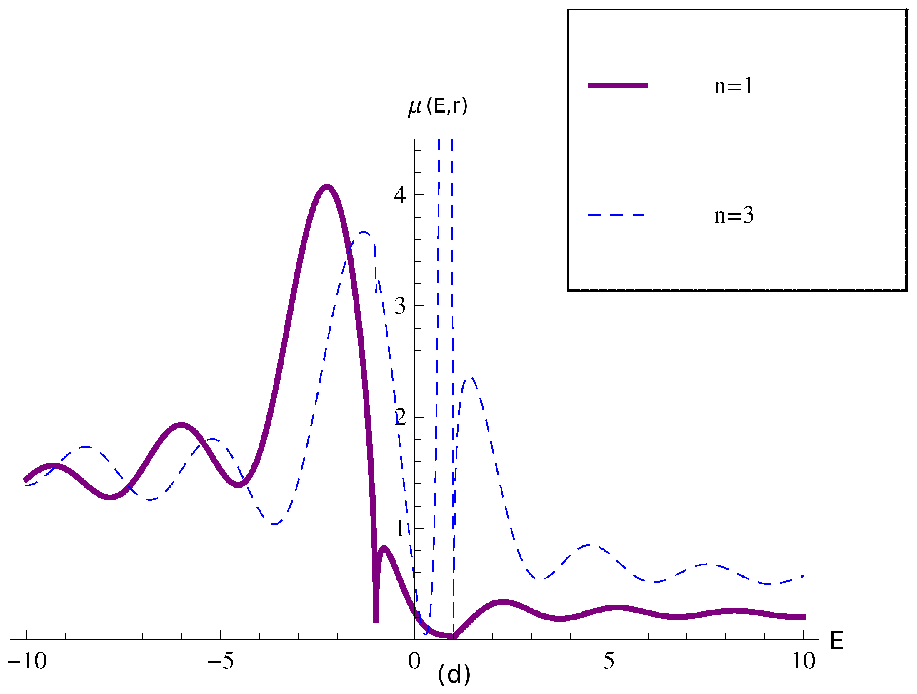,width=0.5\linewidth,clip=}\\
\end{tabular}
\caption{(a)Dependence of LDOS on the distance $r$ from the Coulomb impurity 
is shown for two different values of sae parameter $z=-0.5,0.7$ and a particular value of $E=4$
and with $j=\frac{1}{2},~n=3,~\alpha =0.29$ and $m=1$.
(b) Effect of topology on $r$ dependence of LDOS is shown for sae
parameter $z=-0.5,~E=4$ and system parameters $\alpha=0.29,~n=1,3$ and $m=1$ considering 
contribution coming from the angular momentum channel $j=\frac{1}{2}.$
(c)Energy dependence of LDOS is shown for two different values of sae 
parameter $z= 5,~-0.8$ at a distance $r=1$ from the external Coulomb impurity.
The system parameters used for the plot are $\alpha = 0.29$ and $m=1$ 
and contribution coming from the angular momentum channel $j=\frac{1}{2}$ is considered.
(d) Effect of topology on the energy dependence of LDOS is shown for
sae parameter $z=-0.8$, angular momentum channel $j=\frac{1}{2}$ and system 
parameters $\alpha=-0.29,~n=1,3$ and $m=1$}
\label{fig:8}
\end{figure}

After using the relation (\ref{gbs1}) and matching 
the coefficients of appropriate powers of $r$ at both sides 
in (\ref{gbs1}), we get the following two conditions 

$${(2 \eta )}^{\gamma - \frac{1}{2}}\left ( 1 +\frac{\gamma - \frac{\alpha E}{\eta}}{\nu + \frac{m\alpha}{\eta}}\right ) P_1(k)\sqrt{m + E}~~~~~~~~~~~~~~~~~~~~~~~~~~~~~~~~~~~~~~~~~~~~~$$
\begin{equation} \label{gbs2}
=  C\frac{\pi}{\sin \pi (1+ 2 \gamma)} \left ( \sqrt{m + \frac{i}{d}} e^{i \frac{z}{2} }(P_{+} + R_{+}) + \sqrt{m - \frac{i}{d}} e^{-i \frac{z}{2} }({\bar{P}}_{+} + {\bar{R}}_{+}) \right ) {(2 \eta_1 )}^{\gamma - \frac{1}{2}}
\end{equation}

and 

$${(2 \eta )}^{-\gamma - \frac{1}{2}} \left ( 1 -\frac{\gamma + \frac{\alpha E}{\eta}}{\nu + \frac{m\alpha}{\eta}}\right ) Q_1(k) \sqrt{m + E}~~~~~~~~~~~~~~~~~~~~~~~~~~~~~~~~~~~~~~~~~~~~~$$
\begin{equation} \label{gbs3}
=  -C\frac{\pi}{\sin \pi (1+ 2 \gamma)} \left ( \sqrt{m + \frac{i}{d}} e^{i \frac{z}{2}}(Q_{+} + S_{+}) + 
\sqrt{m - \frac{i}{d}} e^{-i \frac{z}{2}}({\bar{Q}}_{+} + {\bar{S}}_{+}) \right ) {(2 {\eta}_{1} )}^{-\gamma - \frac{1}{2}},
\end{equation}
where $\; z \; $ is the self-adjoint extension parameter. The equations Eq.(\ref{gbs2}) and Eq.(\ref{gbs3}) yield

$$ \frac{(\nu + \frac{m\alpha}{\eta} + \gamma - \frac{\alpha E}{\eta})}{(\nu + \frac{m\alpha}{\eta} - \gamma - \frac{\alpha E}{\eta})} \frac{P_1(k)}{Q_1(k)}  =  - \frac{\sqrt{m + \frac{i}{d}} e^{i \frac{z}{2}}
    (P_{+} + R_{+}) + \sqrt{m - \frac{i}{d}} e^{-i \frac{z}{2}}
    ({\bar{P}}_{+} + {\bar{R}}_{+})}
  {\sqrt{m + \frac{i}{d}} e^{i \frac{z}{2}}
    (Q_{+} + S_{+}) + \sqrt{m - \frac{i}{d}} e^{-i \frac{z}{2}}
    ({\bar{Q}}_{+} + {\bar{S}}_{+})}
   {(2 {\eta}_{1} )}^{2\gamma } {(2 \eta )}^{-2 \gamma}$$

\begin{equation} \label{gbs4}
= - \frac{\chi_1 \cos(\phi_1 + \frac{z}{2})}
  {\chi_2 \cos(\phi_2 + \frac{z}{2})}
   {(2 {\eta}_{1} )}^{2\gamma } {(2 \eta )}^{-2 \gamma },~~~~~~~~~~~~~~~~~~~~~~
\end{equation}
where we have defined  $ \; \sqrt{m + \frac{i}{d}}(P_+ + R_+) \equiv \chi_1 e^{i \phi_1} \; $
and $\; \sqrt{m + \frac{i}{d}}(Q_+ + S_+) \equiv \chi_2 e^{i \phi_2}. \; $ 
Using the relations between the constants given 
before and (\ref{gbs4}), the scattering matrix can now be written as 
\begin{equation} \label{gbs5}
  \mathbf{S} (k) =  
    {(2ik)}^{2 i \frac{\alpha E}{k}}
 \frac{ - \frac{\chi_1 \cos(\phi_1 + \frac{z}{2})}
  {\chi_2 \cos(\phi_2 + \frac{z}{2})}
   {(2 {\eta}_{1} )}^{2\gamma} {(2 \eta)}^{-2 \gamma}
  \frac{ 1 + f_{2}}{1 + f_{1}} f_{1}
 \frac{\Gamma (1 + 2\gamma)}{\Gamma (1 + \gamma + i \frac{\alpha  E}{k})}    
  + f_{2} \frac{\Gamma (1 - 2\gamma)}{\Gamma (1- \gamma + i \frac{\alpha
   E}{k})}}
 { - \frac{\chi_1 \cos(\phi_1 + \frac{z}{2})}
 {\chi_2 \cos(\phi_2 + \frac{z}{2})}
 {(2 {\eta}_{1} )}^{2\gamma } {(2 \eta )}^{-2 \gamma}
 \frac{ 1 + f_{2}}{1 + f_{1}}
 \frac{\Gamma (1 + 2\gamma)}{\Gamma (1 + \gamma -i \frac{\alpha
 E}{k})} e^{-i \pi ( \gamma +i \frac{\alpha E}{k} )}  +
 \frac{\Gamma (1 - 2\gamma)}{\Gamma (1- \gamma -i \frac{\alpha  E}{k})} e^{-i \pi ( - \gamma +i \frac{\alpha E}{k} )}
   },
\end{equation}
where
\begin{equation} \label{gbs6}
f_1 \equiv \frac{\gamma + \frac{i\alpha E}{k}}{\nu - \frac{i m \alpha }{k} },
\quad  f_2 \equiv \frac{-\gamma + \frac{i\alpha E}{k}}{\nu - \frac{i m \alpha }{k} }.    
\end{equation}
The expression in (\ref{gbs5}) gives the $\mathbf{S}$ matrix for gapped graphene
for the parameter range $\; 0 < \gamma < \frac{1}{2} \; $. For this range
of $\gamma$, the appropriate boundary conditions for which the
Hamiltonian  is self-adjoint and the corresponding time 
evolution is unitary, requires the introduction of an additional real 
self-adjoint extension parameter $z$, which labels the allowed boundary 
conditions. The phase shifts and the $S$ matrix 
depend explicitly on $z$. For each value of $z$ (mod $2 \pi$), we have 
an inequivalent set of the scattering data. Practically realizable 
value of $z$ should be determined empirically as it cannot be determined 
analytically.

In Figure $7(a)$ we have shown the energy dependence of scattering phase shifts 
for three different values of the self adjoint extension parameter $z$. We
can see from the Figure that the region $|E|<|m|$ is characterized by the sharp discontinuous 
oscillations which indicate the appearance of discrete bound states. In the 
other region where $|E|>|m|$, scattering phase shifts behave qualitatively
in the same manner for different values of $z$ but they are clearly 
distinguishable from each other. In Figure $7(b)$ we have observed the effect 
of topology on the energy dependence of scattering phase shifts for a 
particular self adjoint extension parameter. It should be noted that 
the Figure shows the effect of topology considering only two values 
of $n (n=~1,3)$ because the value of $\gamma$ remains in the region  
$\; 0 < \gamma < \frac{1}{2} \; $ for only those two values. 
During the analysis we have always restricted the 
obtained results to the  parameter range $0<\gamma<\frac{1}{2}$ through 
the appropriate choice of system parameters. In the plots we have assumed that $d=1$.

In Figures $8(a)$ and $8(b)$ we have shown the dependence of LDOS on the distance $r$ from 
the external Coulomb impurity placed at the apex of the gapped graphene cone where the energy
is fixed at a value $E=4m$. We have used
the following expression given in equation Eq.(\ref{ldos}) for LDOS during the plotting.
\begin{eqnarray}
\label{ldos}
\mu(E,r)=\frac{4}{\pi \hbar v_F}\sum_{j}|\Psi^{(j)}(k,r)|^2.
\end{eqnarray}
For the numerical calculation we have used the Equations (\ref{sub1.1}),(\ref{sub1.2}),
(\ref{scat1}),(\ref{scat2}) and (\ref{gbs4}). In Figures $8(c)$ and $8(d)$ we have
plotted the energy dependence of LDOS at a distance close to the charge impurity $(r=1)$.
From these Figures we can observe that LDOS depend on the values of self adjoint 
extension parameter $z$ and also on the topology of the system. Therefore measurement of LDOS 
using scanning tunneling microscopy can give us information about the self adjoint extension 
parameter and the topology of the system.

\section{Properties of gapped graphene cone with supercritical Coulomb charge}

In the supercritical region the radial part of the Dirac equation 
obeyed by the gapped graphene cone appears to be the same as Eq.(\ref {sub1.3}).
The only difference is that in this region $\gamma$ is always imaginary as the 
Coulomb potential strength $\alpha$ exceeds the value of $\nu$. We denote 
$\gamma=i\lambda=\sqrt{\alpha^2 - \nu^2}$.
Then from Eq.(\ref{sub1.3}) we have
\begin{equation} 
\label{g2}
 \rho \frac{d F}{d \rho} + (i\lambda -\frac {\alpha E}{\eta}) F - (\nu +\frac{m\alpha}{\eta}) G = 0,
\end{equation}
and
\begin{equation} \label{g3}
 \rho \frac{d G}{d \rho} + (i\lambda -\rho +\frac {\alpha E}{\eta}) G + (-\nu + \frac{m\alpha}{\eta}) F = 0.~~~~~~~~~
\end{equation} 
Substituting the expression of $G$ from Eq.(\ref{g2}) in Eq.(\ref{g3}) we have
\begin{equation}\label{g4}
 \rho F^{\prime\prime} + (1+2 i \lambda-\rho)F^{\prime} - (i \lambda -\frac{\alpha E}{\eta})F=0.
\end{equation}
Solving the differential equations we can obtain the low energy eigenstates of the gapped graphene 
cone using two different boundary conditions. In the next section 
we shall consider a regularized Coulomb potential and obtain the 
quasibound state energy spectrum and then we shall observe how 
the nonzero mass and cutoff parameter affects the critical charge  
of the system. We shall repeat the same calculations with the zigzag edge 
boundary condition also.

\subsection{Regularized Coulomb Potential}

In order to observe the supercritical effect of external Coulomb charge on the gapped graphene cone 
we shall first consider a regularized Coulomb potential because in that case we are allowed to extend the 
bound states  until the negative continuum $E=-m$ is reached\cite{castro2,gamayun}. The regularization of the Coulomb 
potential given by 
\begin{equation}\label{g13}
  V(r) = \begin{cases}
           - \alpha/r, & r > a\\
           - \alpha/a, & r \leq a
         \end{cases},  
\end{equation}
where the Coulomb charge is placed at the apex of the gapped graphene cone, 
$a$ is the minimum distance of the Dirac electron from the apex and it is 
of the order of the lattice parameter. The Dirac equation for gapped graphene cone 
is solved for these two different regions.

Let us first consider the region $r\leq a$. In this region the Dirac equation is given by 
\begin{eqnarray}
\label{g14}
\left( 
\begin{array}{cc}
(E-m+\frac{\alpha}{a}) & -\{ \partial_r +(\nu +\frac{1}{2})\frac{1}{r}\}  \\
\{\partial_r -(\nu - \frac{1}{2})\frac{1}{r}\} & (E+m+\frac{\alpha}{a})
\end{array}
\right)\left( \begin{array}{c}
 {P_1^{(j)}(r)} \\
 {Q_1^{(j)}(r)} \\
\end{array} \right)=0.
\end{eqnarray}
Eq. (\ref{g14}) gives the following two coupled first order differential equations.
\begin{equation}\label{g15}
 -Q_1^{\prime (j)}(r)-\frac{(\nu + \frac{1}{2})}{r}Q_1^{(j)}(r) + \left(E - m + \frac{\alpha}{a}\right) P_1^{(j)}(r)=0
\end{equation}
 and
\begin{equation}\label{g16}
 P_1^{\prime (j)}(r)-\frac{(\nu - \frac{1}{2})}{r}P_1^{(j)}(r) + \left(E + m + \frac{\alpha}{a}\right) Q_1^{(j)}(r)=0. 
\end{equation}
Substituting the expression of $Q_1^{(j)}(r)$ from Equation (\ref{g16}) in Equation (\ref{g15})
we have
\begin{equation}\label{g17}
 r^2 P_1^{\prime \prime (j)}(r) + r P_1^{\prime (j)}(r)+ \left[\left\{\left(E + \frac{\alpha}{a}\right)^2 - m^2\right\}r^2 - \left(\nu -\frac{1}{2}\right)^2\right]=0.
\end{equation}
A general solution of this Bessel equation is given by
\begin{equation}\label{g18}
 P_1^{(j)}(r) = A_1 J_{|\nu - \frac{1}{2}|} \left(r\sqrt{(E + \frac{\alpha}{a})^2 - m^2}\right).
\end{equation}
Using Eq. (\ref{g18}) we obtain
\begin{equation}\label{g19}
 Q_1^{(j)}(r) = \sqrt{\frac{(E + \frac{\alpha}{a} -m)}{(E + \frac{\alpha}{a} + m)}} A_1 J_{|\nu + \frac{1}{2}|} \left(r\sqrt{\left(E + \frac{\alpha}{a}\right)^2 - m^2}\right).
\end{equation}
Now we consider the region $r > a$. In this case the coupled 1st order differential equations
and the second order differential equations obeyed by the Dirac fermions will be the same as Equations (\ref{g2}), 
(\ref{g3}) and (\ref{g4}). Then using the regularity condition at infinity we have \cite{stegun}
\begin{equation}\label{g20}
 F = U\left(i\lambda-\frac{E\alpha}{\eta},1+2i\lambda,\rho \right)
\end{equation}
and
\begin{equation}\label{g21}
 G = \left(\frac{m\alpha}{\eta} - \nu \right) U\left(1+i\lambda-\frac{E\alpha}{\eta},1+2i\lambda,\rho\right).
\end{equation}
Therefore the upper and lower components of the Dirac wave function will be given by
\begin{equation}\label{g22}
 P_2^{(j)}(r) = \sqrt{m+E} e^{-\eta r} (2 \eta r)^{(i\lambda - \frac{1}{2})} \left[U\left(i\lambda-\frac{E\alpha}{\eta},1+2i\lambda,\rho\right) + \left(\frac{m\alpha}{\eta} - \nu \right) U\left(1+i\lambda-\frac{E\alpha}{\eta},1+2i\lambda,\rho \right)\right]
\end{equation}
and 
\begin{equation}\label{g23}
 Q_2^{(j)}(r) = \sqrt{m-E} e^{-\eta r} (2 \eta r)^{(i\lambda - \frac{1}{2})} \left[U\left(i\lambda-\frac{E\alpha}{\eta},1+2i\lambda,\rho \right) - \left(\frac{m\alpha}{\eta} - \nu \right) U\left(1+i\lambda-\frac{E\alpha}{\eta},1+2i\lambda,\rho \right)\right].
\end{equation}                                                                                                                                                                                                                                                                                                                       
To determine the bound states we use the continuity condition of the wave function at $r=a$. It is given by
\begin{equation}\label{g24}
 \frac{P_1^{(j)}(r)}{Q_1^{(j)}(r)}|_{r=a} = \frac{P_2^{(j)}(r)}{Q_2^{(j)}(r)}|_{r=a}.
\end{equation}
\begin{figure}
\centering
\begin{tabular}{cc}
\epsfig{file=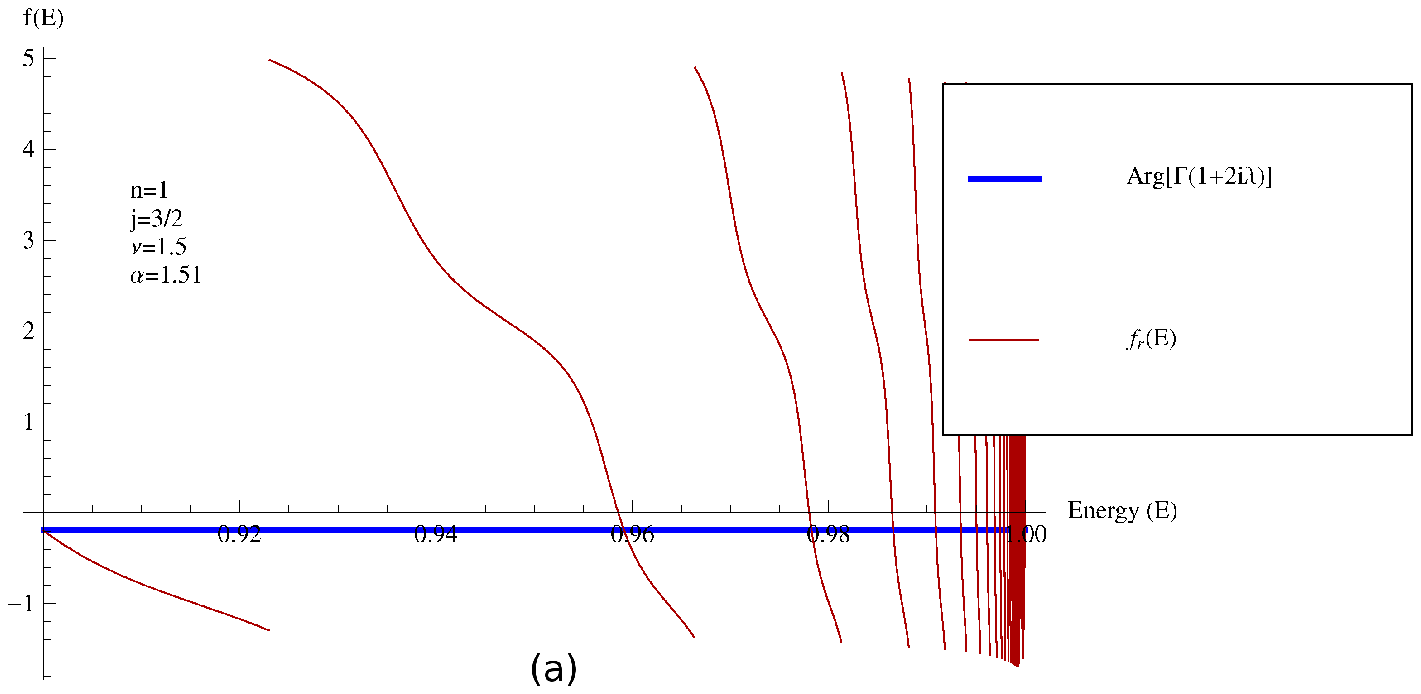,width=0.5\linewidth,clip=} &
\epsfig{file=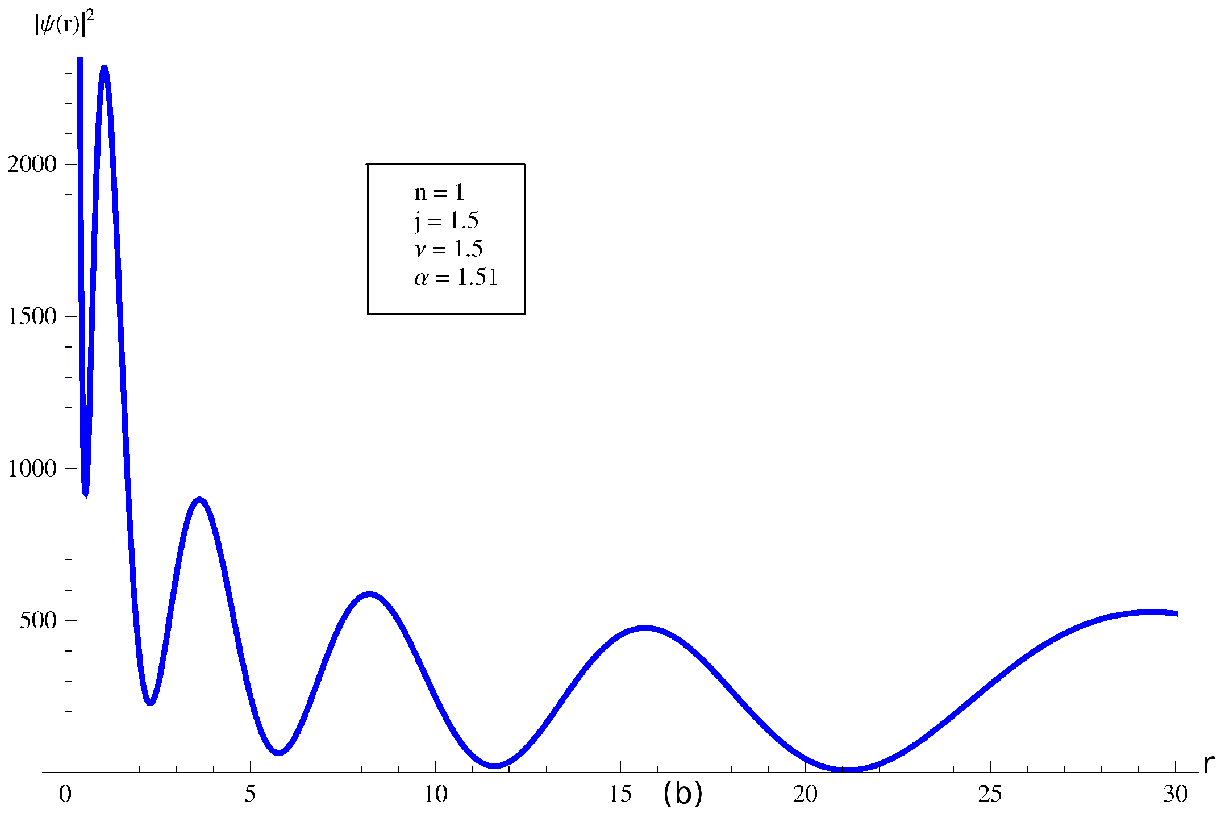,width=0.5\linewidth,clip=}\\
\end{tabular}
\caption{(a)Bound state energy spectrum with regularized potential is shown. 
Here the blue line represents $\mbox{Arg}[\Gamma(1+2i\lambda)]$
and the dark red line represents RHS of Equation (\ref{g27})).
(b)Dependence of $|\Psi(r)|^2$ on the distance $r$ from the charge impurity 
placed at the apex of the gapped graphene cone is shown
for a particular energy $E=0.96m$ obtained from the plot of the
bound state energy spectrum.
}
\label{fig:3}
\end{figure}
Using this condition given in Eq. (\ref{g24}) we have
\begin{equation}\label{g25}
 \frac{U\left(i\lambda-\frac{E\alpha}{\eta},1+2i\lambda,2\eta a \right)}{\left(\frac{m\alpha}{\eta} - \nu \right) U\left(1+i\lambda-\frac{E\alpha}{\eta},1+2i\lambda,2\eta a \right)} = -\frac{\mu + 1}{\mu - 1}
\end{equation}
where
\begin{equation}\label{g26}
\mu = \sqrt{\frac{(m+E)(E + \frac{\alpha}{a} - m)}{(m-E)(E + \frac{\alpha}{a} + m)}} \frac{J_{|\nu + \frac{1}{2}|}(\sqrt{(Ea + \alpha)^2 - m^2 a^2})}{J_{|\nu - \frac{1}{2}|}(\sqrt{(Ea + \alpha)^2 - m^2 a^2})}.
\end{equation}
When $a\longrightarrow 0$, the continuity condition given in Eq. (\ref{g24}) will be satisfied when
\begin{eqnarray}\label{g27}
f(E)\equiv\frac{\Gamma(1+i\lambda-\frac{E\alpha}{\eta})}{\Gamma(1-i\lambda-\frac{E\alpha}{\eta})} e^{2i\lambda \mbox{ln}(2\eta a)}\left[\frac{\nu - \frac{\alpha}{\eta}(m-E)+i\lambda}{\nu-\frac{\alpha}{\eta}(m-E)-i\lambda}\right]\left[\frac{\nu - \alpha \frac{J_{|\nu + \frac{1}{2}|}(\alpha)}{J_{|\nu - \frac{1}{2}|}(\alpha)}-i\lambda}{\nu-\alpha \frac{J_{|\nu+\frac{1}{2}|}(\alpha)}{J_{|\nu-\frac{1}{2}|}(\alpha)}+i\lambda}\right]=\frac{\Gamma(1+2i\lambda)}{\Gamma(1-2i\lambda)}~~~~~~~~~~~~~~~~~~~~~~~~~~~~~~~~~~\nonumber\\
\mbox{or}f(E)\equiv f_{r}(E)=\mbox{Arg}[\Gamma(1+2i\lambda)]~~~~~~~~~~~~~~~~~~~~~~~~~~~~~~~~~~~~~~~~~~~~~~~~~~~~~~~~~~~~~~~~~~~~~~~~~~~~~~~~~~~~~~~~~~~~~~~~~~~~~~~~~~~~~~~~~~~\nonumber\\
\end{eqnarray}
where p is a positive integer and
$$f_{r}(E)= \mbox{Arg}\left[\Gamma\left(1+i\lambda-\frac{E\alpha}{\eta}\right)\right]+\mbox{Arg}\left[\nu-\frac{\alpha}{\eta}(m-E)+i\lambda\right]+\lambda \mbox{ln}(2\eta a)+ \mbox{Arg}\left[\nu- \alpha \frac{J_{|\nu + \frac{1}{2}|}(\alpha)}{J_{|\nu - \frac{1}{2}|}(\alpha)}-i\lambda\right] +p\pi. $$
Eq. (\ref{g27}) gives the bound state energy spectrum 
in presence of a regularized Coulomb potential for all possible situations.

Now we concentrate to the physically 
interesting cases where $|E|\gg m$ and the Coulomb potential is near to its 
critical value i.e $\alpha\approx\alpha_c=|\nu_{\mbox{min}}|$ for a 
particular $n$. Then from the Equation (\ref{g27}) we have 
\begin{eqnarray}\label{g28}
\mbox{ln}\left(-2iE_p \sqrt{1-\frac{m^2}{E_p^2}}a\right)=2 \psi(1)+\frac{J_{|\nu - \frac{1}{2}|}(\nu)}{\nu[J_{|\nu-\frac{1}{2}|}(\nu)-J_{|\nu + \frac{1}{2}|}(\nu)]}-\psi\left(1-\frac{i\nu}{\sqrt(1-\frac{m^2}{E_p^2})}\right)-\frac{1}{\nu\left[1+i\sqrt{\frac{1-\frac{m}{E_p}}{1+\frac{m}{E_p}}}\right]}-\frac{p\pi}{\lambda}.~~
\end{eqnarray}
Here $\psi(x)=\frac{\Gamma^{\prime}(x)}{\Gamma(x)}$ and $\eta=-i\sqrt{E_p^2-m^2}$. 
For $\frac{m}{E}<<1$ we have up to the terms of order $\frac{m^2}{E^2}$,
\begin{eqnarray}\label{g29}
E_p - \frac{m^2}{2E_p}=\frac{1}{2a} \mbox{exp}\left[\frac{J_{|\nu-\frac{1}{2}|}(\nu)}{\nu(J_{|\nu-\frac{1}{2}|}(\nu)-J_{|\nu+\frac{1}{2}|}(\nu))}+2\psi(1)-\psi(1-i\nu)-\frac{(1-i)}{2\nu}-\frac{p\pi}{\lambda} + \frac{i\pi}{2}\right]\nonumber\\
\left[1-\frac{m}{E_p}\left(\frac{1}{2\nu}\right)+\frac{m^2}{E_p^2}\left\{\frac{i\nu}{2}\psi^{\prime}(1-i\nu)-\frac{i}{4\nu}\right\}\right].
\end{eqnarray}
Thus we can see the effect of the nonzero mass on the bound state energy spectrum.

The mass affects the critical charge of the system. For a regularized Coulomb
potential the bound states can dive into negative energies. Here the critical charge
refers to that value of Coulomb potential for which $E=-m$. Then for the region 
near critical potential we have
\begin{eqnarray}\label{g30}
 \alpha_c = \nu + \frac{\pi^2}{2\nu \mbox{log}^2[2m\nu C a]}
\end{eqnarray}
where 
\begin{equation}\label{g31}
 C= \mbox{exp}\left[-2\psi(1)-\frac{J_{|\nu-\frac{1}{2}|}(\nu)}{\nu(J_{|\nu-\frac{1}{2}|}(\nu)-J_{|\nu+\frac{1}{2}|}(\nu))}\right].
\end{equation}
From Equation (\ref{g30}) we can see when $ma\rightarrow0, \alpha_c \approx \nu$ which agrees with the result
obtained for massless case \cite{critical}. The dependence of critical charge on the nonzero mass and cutoff parameter are shown 
in Fig.(\ref{fig:5}) for different opening angles of the gapped graphene cone.
From the Fig.(\ref{fig:5}) it is clear that the topology affects the 
critical charge of the system and the nature of their dependence 
on the product of mass and cutoff parameter of the system remains almost same.

\subsection{Zigzag edge boundary condition}
To find out the energies of the stationary states formed 
in the supercritical region now we use the zigzag edge boundary 
condition $\Psi_B^j(a)=0$, where $a$ is a distance from the apex, of the 
order of the lattice scale in graphene. In order to proceed we shall first
solve Eq.(\ref{g4}) and obtain two linearly independent solutions $F_1(\rho)$ and $F_2(\rho)$
which are regular at $\rho=0$. They are given by \cite{stegun}
\begin{equation}\label{g5.1}
 F_1 (\rho)= A_1 M(i\lambda -\frac{E\alpha}{\eta}, 1+2 i\lambda, \rho) 
\end{equation}
and
\begin{equation}\label{g5.2}
 F_2 (\rho) = A_2 \rho^{-2 i \lambda}M(-i\lambda-\frac{E\alpha}{\eta}, 1-2 i\lambda, \rho).
\end{equation}
Then from Eq.(\ref{g2}) we have
\begin{equation}\label{g6.1}
 G_1(\rho)= \frac{(i\lambda-\frac{\alpha E}{\eta})}{(\nu +\frac{m\alpha}{\eta})}A_1 M(1 + i\lambda -\frac{E\alpha}{\eta}, 1+2 i \lambda, \rho)
\end{equation}
and 
\begin{equation}\label{g6.2}
 G_2(\rho)=-\frac{(i \lambda + \frac{\alpha E}{\eta})}{(\nu +\frac{m\alpha}{\eta})}A_2 \rho^{-2 i\lambda} M(1-i\lambda-\frac{E\alpha}{\eta}, 1-2 i \lambda, \rho).
\end{equation}
The solution satisfying the zigzag edge boundary condition can be given by
\begin{equation}\label{g7}
\Psi_B^j(r)=\sqrt{m-E}e^{-\frac{\rho}{2}} {\rho}^{\gamma - \frac{1}{2}}[\{F_1(a)-G_1(a)\}\{F_2(r)-G_2(r)\}-\{F_1(r)-G_1(r)\}\{F_2(a)-G_2(a)\}]
\end{equation}
The square integrability condition of the wave function indicates that 
as $\rho\rightarrow\infty$ the diverging part of the wave function 
must vanish. Therefore we have
\begin{eqnarray}\label{g8}
\sqrt{m-E}e^{\frac{\rho}{2}} {\rho}^{(-\frac{E\alpha}{\eta} - \frac{3}{2})}A_1 A_2 \left[\frac{\Gamma(1-2i\lambda)}{\Gamma(-i\lambda-\frac{E\alpha}{\eta})}\left\{1-\frac{(i\lambda-\frac{E\alpha}{\eta})}{(\nu + \frac{m\alpha}{\eta})}\right\}-\frac{\Gamma(1+2i\lambda)}{\Gamma(i\lambda-\frac{E\alpha}{\eta})}\left\{1-\frac{(-i\lambda-\frac{E\alpha}{\eta})}{(\nu + \frac{m\alpha}{\eta})}\right\}e^{-2i\lambda \mbox{ln}(2\eta a)}\right]\nonumber\\
\left[1-\frac{(\rho-1-\frac{2 E \alpha}{\eta})}{(\nu + \frac{m\alpha}{\eta})}\right]=0.~~~~
\end{eqnarray}
This gives the condition
\begin{eqnarray}\label{g9}
\frac{\Gamma(1-2i\lambda)}{\Gamma(-i\lambda-\frac{E\alpha}{\eta})}\left\{1-\frac{(i\lambda-\frac{E\alpha}{\eta})}{(\nu + \frac{m\alpha}{\eta})}\right\}=\frac{\Gamma(1+2i\lambda)}{\Gamma(i\lambda-\frac{E\alpha}{\eta})}\left\{1-\frac{(-i\lambda-\frac{E\alpha}{\eta})}{(\nu + \frac{m\alpha}{\eta})}\right\}e^{-2i\lambda \mbox{ln}(2\eta a)}\nonumber\\
\mbox{or}~~~f(E)\equiv \frac{\Gamma(i\lambda-\frac{E\alpha}{\eta})}{\Gamma(-i\lambda-\frac{E\alpha}{\eta})} e^{2i\lambda \mbox{ln}(2\eta a)}\left[\frac{\nu + \frac{\alpha}{\eta}(m+E)-i\lambda}{\nu+\frac{\alpha}{\eta}(m+E)+i\lambda}\right]=\frac{\Gamma(1+2i\lambda)}{\Gamma(1-2i\lambda)}~~~~~~~~~~~~\nonumber\\
\mbox{or}~~~f(E)\equiv f_{z}(E)=\mbox{Arg}[\Gamma(1+2i\lambda)]~~~~~~~~~~~~~~~~~~~~~~~~~~~~~~~~~~~~~~~~~~~~~~~~~~~~~~~~~~~~
\end{eqnarray}
where p is a positive integer and
$$f_{z}(E)=\mbox{Arg}\left[\Gamma\left(i\lambda-\frac{E\alpha}{\eta}\right)\right]+\mbox{Arg}\left[\nu+\frac{\alpha}{\eta}(m+E)-i\lambda\right]+\lambda \mbox{ln}(2\eta a)+ p\pi.$$
This is the condition directly obtained 
from our analysis and obeyed by the bound state energy spectrum for 
all possible situations. 
\begin{figure}
\centering
\begin{tabular}{cc}
\epsfig{file=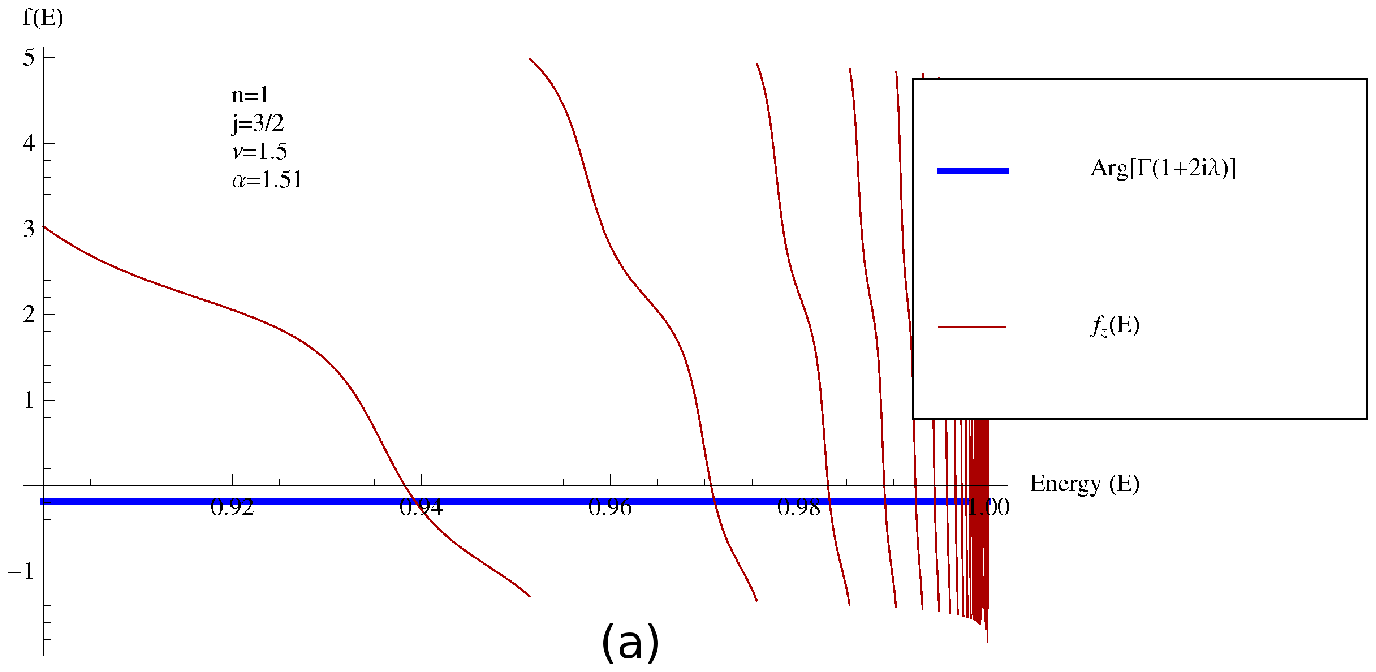,width=0.5\linewidth,clip=} &
\epsfig{file=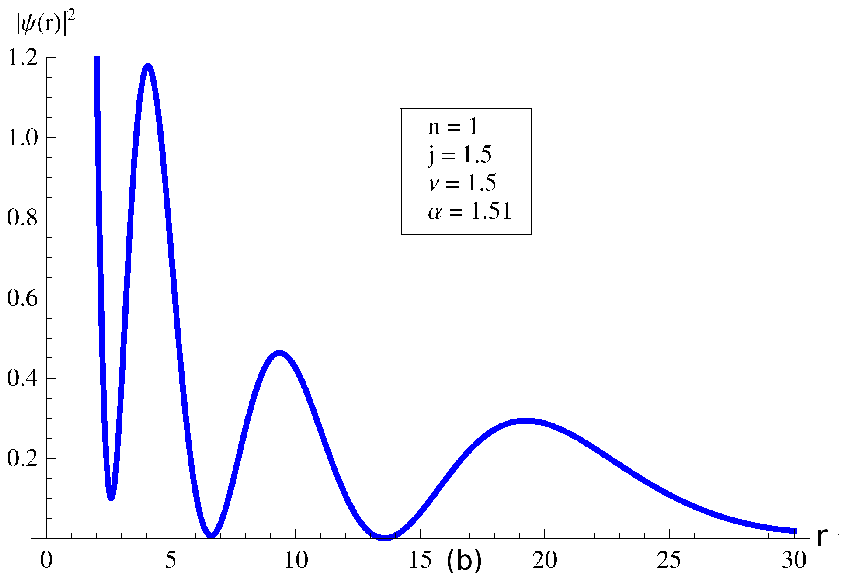,width=0.5\linewidth,clip=}\\
\end{tabular}
\caption{(a)Bound state energy spectrum with zigzag edge boundary condition is shown. 
Here the blue line represents $\mbox{Arg}[\Gamma(1+2i\lambda)]$
and the red line represents RHS of Equation (\ref{g9})).
(b)Dependence of $|\Psi(r)|^2$ on the distance $r$ from the charge impurity 
placed at the apex of the gapped graphene cone is shown
for a particular energy $E=0.94m$ obtained from the plot of the
bound state energy spectrum.
}
\label{fig:2}
\end{figure}

We can compare this bound state energy spectrum for zigzag edge boundary with the spectrum obtained from the regularized potential case 
and observe how the two different boundary conditions affect the spectrum. 
\begin{figure}
\centering
\begin{tabular}{cc}
\epsfig{file=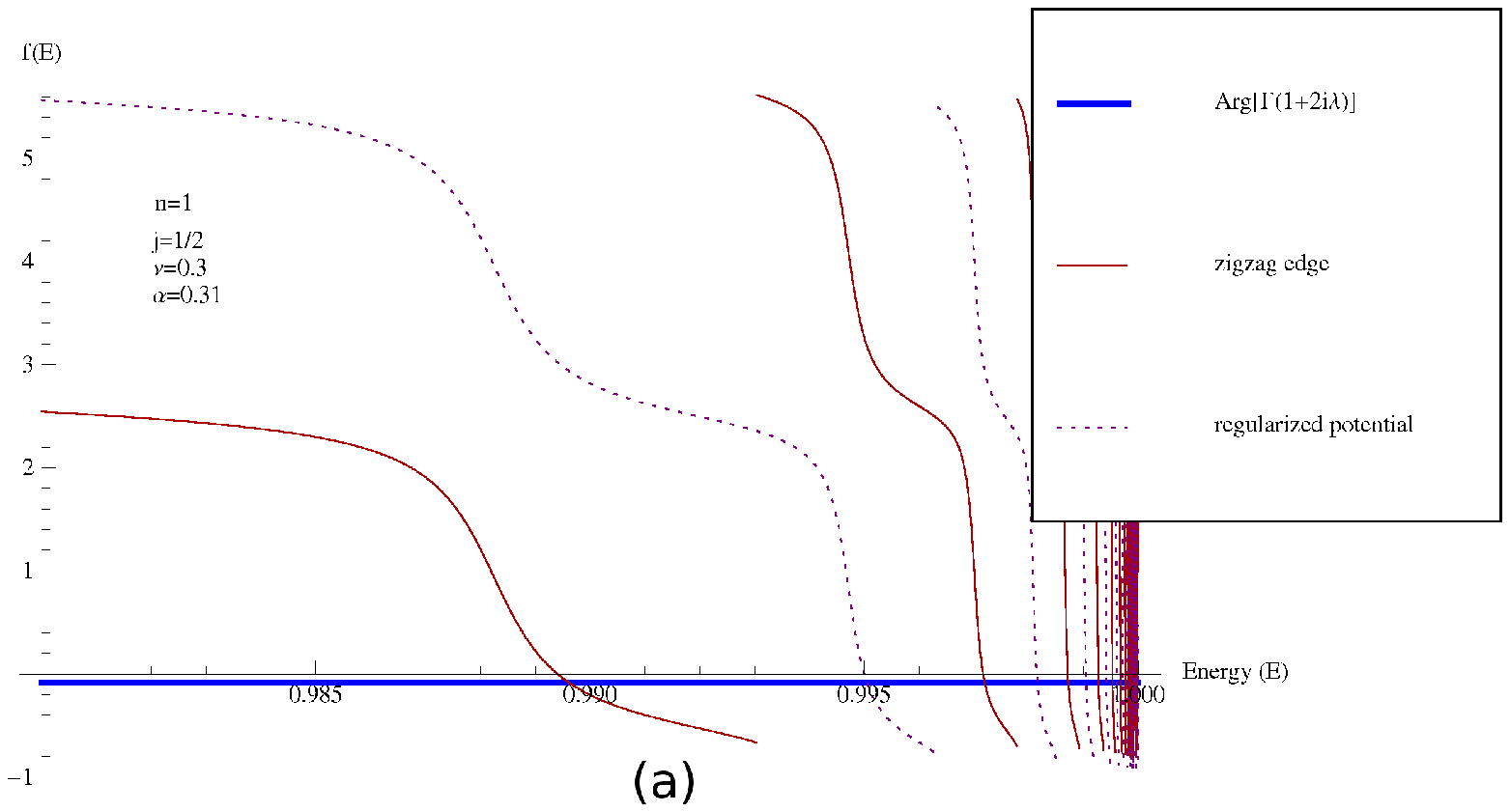,width=0.5\linewidth,clip=} &
\epsfig{file=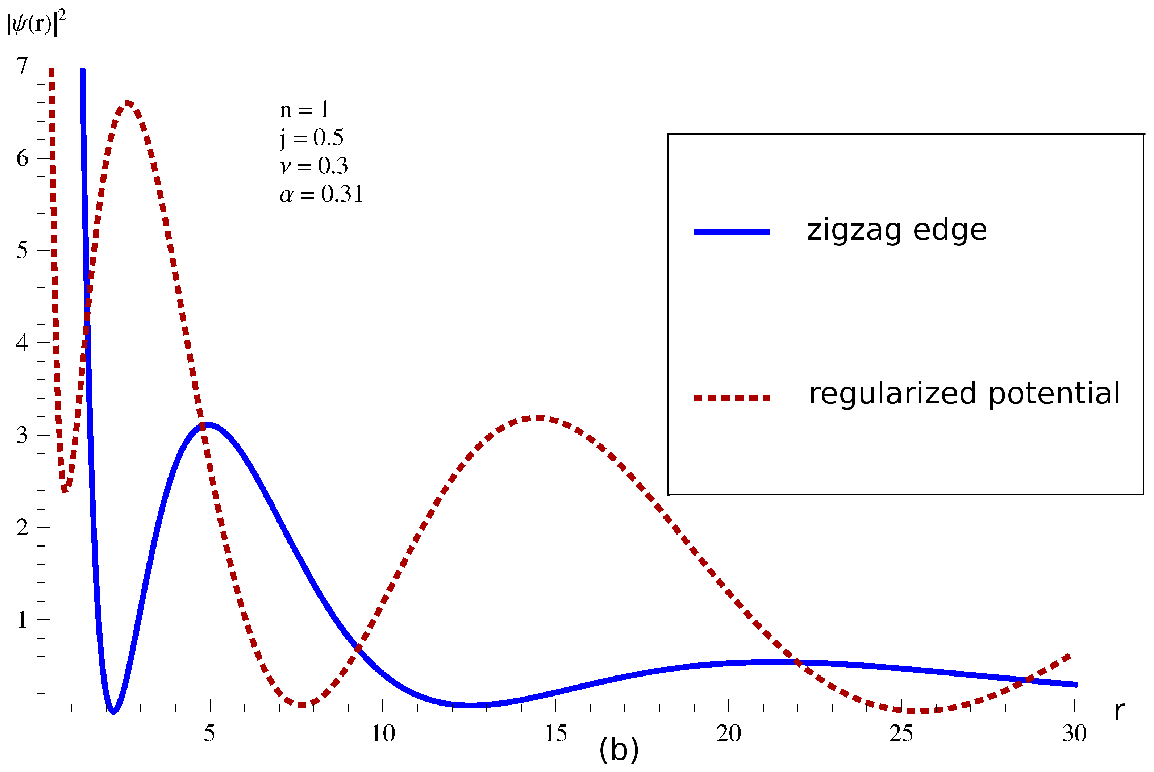,width=0.5\linewidth,clip=}\\
\end{tabular}
\caption{(a)Bound state energy spectrum with zigzag edge boundary condition and regularized Coulomb potential. 
Here the blue line represents $\mbox{Arg}[\Gamma(1+2i\lambda)]$ and the dashed and the solid line represents 
RHS of Equations (\ref{g9}) and (\ref{g27}) respectively.
(b)Dependence of $|\Psi(r)|^2$ on the distance $r$ from the charge impurity 
placed at the apex of the gapped graphene cone is shown 
for both the zigzag edge boundary condition and regularized Coulomb potential.
The values of energy are obtained from the bound state energy spectrum. 
From Fig.(a) we can see that for zigzag edge boundary condition a possible
bound state energy is $E=0.9895m$ and for regularized potential a
possible energy is $E=0.995m$. Here we have given the probability amplitude 
plots for these two particular energies.
}
\label{fig:4}
\end{figure}
From Fig.(\ref{fig:4}) we can observe that the bound state energy spectra
and the probability amplitude of the wavefunction gets affected by the boundary 
conditions though their nature remains same.

Now we again concentrate to some physically 
interesting cases with some approximations. First we consider 
the case where $|E|\gg m$ and the Coulomb potential is near to its 
critical value i.e $\alpha\approx\alpha_c=|\nu_{\mbox{min}}|$ for a 
particular $n$. Then from the Equation (\ref{g9}) we have 
\begin{eqnarray}\label{g10}
 \mbox{ln}\left(-2iE_p \sqrt{1-\frac{m^2}{E_p^2}}a\right)=2 \psi(1)-\psi\left(-\frac{i\nu}{\sqrt(1-\frac{m^2}{E_p^2})}\right)+\frac{1}{\nu\left[1+i\sqrt{\frac{1+\frac{m}{E_p}}{1-\frac{m}{E_p}}}\right]}-\frac{p\pi}{\lambda}.
\end{eqnarray}
Like the previous section here also $\psi(x)=\frac{\Gamma^{\prime}(x)}{\Gamma(x)}$ and $\eta=-i\sqrt{E_p^2-m^2}$. 
For $\frac{m}{E}<<1$ we have up to the terms of order $\frac{m^2}{E^2}$,
\begin{eqnarray}\label{g11}
 E_p - \frac{m^2}{2E_p}=\frac{1}{2a} \mbox{exp}\left[\left\{2\psi(1)-\psi(-i\nu)+\frac{(1-i)}{2\nu}-\frac{p\pi}{\lambda}\right\}-\frac{m}{E_p}\left(\frac{1}{2\nu}\right)+\frac{m^2}{E_p^2}\left\{\frac{i\nu}{2}\psi^{\prime}(-i\nu)+\frac{i}{4\nu}\right\}\right]\nonumber\\
\mbox{or}~~~E_p - \frac{m^2}{2E_p}=\frac{1}{2a} \mbox{exp}\left[2\psi(1)-\psi(-i\nu)+\frac{(1-i)}{2\nu}-\frac{p\pi}{\lambda}\right]\left[1-\frac{m}{E_p}\left(\frac{1}{2\nu}\right)+\frac{m^2}{E_p^2}\left\{\frac{i\nu}{2}\psi^{\prime}(-i\nu)+\frac{i}{4\nu}\right\}\right].
\end{eqnarray}
Thus we can see the effect of the nonzero mass on the bound state energy spectrum. 

Like the critical charge obtained for regularized Coulomb potential, here also with zigzag edge boundary condition we can see 
that the mass affects the critical charge of the system. Proceeding as before in this case we have 
\begin{eqnarray}\label{g12}
 \alpha_c = \nu + \frac{\pi^2}{2\nu \mbox{log}^2[2m\nu a \mbox{exp}(-2\psi(1))]}.
\end{eqnarray}
Here also from Equation (\ref{g12}) we can see when $ma\rightarrow0, \alpha_c \approx \nu$ which agrees with the result
obtained for massless case \cite{critical}. The dependence of critical charge on the nonzero mass and cutoff parameter has been
shown for different opening angles of the gapped graphene cone. Also we have compared 
the dependence for two different boundary conditions.

\begin{figure}
\centering
\includegraphics[bb=160 14 140 220]{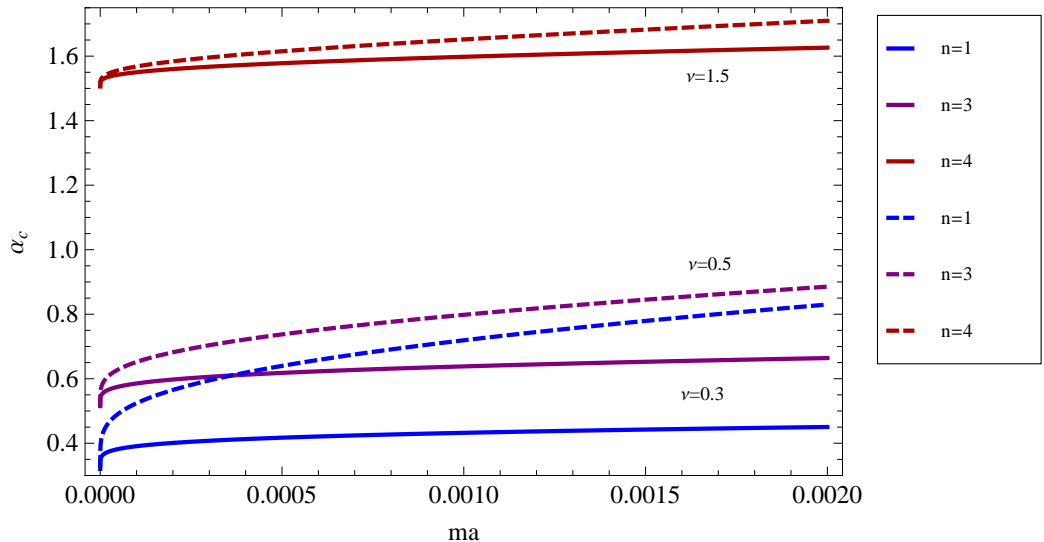}
\caption{
Dependence of critical charge on the nonzero mass and cutoff parameter are shown 
for both zigzag edge boundary condition and regularized Coulomb potential for different opening
angles of the gapped graphene cone. The dotted lines show the dependence for zigzag edge
boundary condition and the solid lines show the dependence for regularized Coulomb potential
}
\label{fig:5}
\end{figure}
From Fig.(\ref{fig:5}) we can see that for a zigzag edge boundary condition 
the critical charge vary with $ma$ more rapidly than it varies for a 
regularized Coulomb potential. Thus the two different boundary conditions 
affect the dependence of critical charge of a gapped graphene cone on $ma$.

\newpage
\section{Conclusion}

In this paper we have described the low energy dynamics of massive
 Dirac fermions in a gapped graphene cone in the presence of an 
external Coulomb charge impurity. The graphene cone can be equivalently
 described as a graphene plane together with a flux tube whose gauge 
potential is chosen to produce the required holonomies. The given system 
thus consists of a gapped graphene plane together with a combination of a 
Coulomb charge impurity and a flux tube passing through it. The strength 
of the Coulomb charge can be sub or supercritical.

The combination of this topological defect as well as the charge impurity 
results in short distance interactions, the effect of which cannot be 
incorporated as dynamical terms in the Dirac equation, valid in the low energy 
limit. For a sub critical charge impurity, we show that the effect of these 
interactions can be modelled through appropriate choice of boundary conditions, 
which are determined by imposing the requirement of a unitary time evolution. 
While there is a very large class of allowed boundary conditions, it turns out 
that they can be labelled by a single real parameter. It is this parameter through 
which the effect of the short range interactions enter in the analysis presented here. 
This parameter cannot be determined from theory alone. However, we have shown that 
the observables such as LDOS, scattering phase shifts and bound state energies depend 
explicitly on this parameter. As mentioned before, a similar situation arose in the 
context of Dirac fermions in a plane in the presence of a cosmic string. These two 
situations are not identical, but similar quantum subtleties arise in both contexts 
and the case of the gapped graphene considered here is more amenable to empirical analysis. 

The supercritical regime of the external Coulomb charge is characterized by quantum
instabilities. Here we have analyzed the effect of topological defects in the supercritical 
dynamics of gapped graphene. This problem has been analyzed with a regularized Coulomb potential 
as well as with a zigzag boundary condition. We have shown that the quasibound state spectra and the probability amplitude
depend explicitly on the number of sectors removed from a planar 
graphene to form the cone. In addition, the running of the critical charge as a function of the product of the Dirac mass $m$ and the cutoff parameter $a$ has been obtained. Though the nature of the dependence is 
similar for both the regularized Coulomb potential and the zigzag boundary condition
but in the latter case the critical charge increases more rapidly with $ma$ than the former case.  

It would be interesting to study the effect of sample topology on the electron-electron interactions 
in graphene and the associated gap equation, which is currently under investigation.

\end{document}